\documentclass[%
 reprint,
 amsmath,amssymb,
 aps,
]{revtex4-2}

\usepackage{graphicx}
\usepackage{subcaption}
\usepackage{dcolumn}
\usepackage{tabularx}
\usepackage{multirow}
\usepackage{bm}
\usepackage{amsfonts,color} 
\usepackage[table,xcdraw]{xcolor}
\usepackage[english]{babel}
\usepackage{lipsum} 
\usepackage{array}

\definecolor{Gray}{gray}{0.9}

\newcolumntype{Y}{>{\centering\arraybackslash}X}
\newcolumntype{Z}{>{\raggedright\arraybackslash}X}

\newcolumntype{P}[1]{>{\raggedright\arraybackslash}p{#1}}

\begin{document}

\preprint{APS/123-QED}

\title{
Making expert processes visible: how and why theorists use assumptions and analogies in their research
}

\author{Mike Verostek}
  \email{mveroste@ur.rochester.edu}
\affiliation{
 Department of Physics and Astronomy, University of Rochester, Rochester, New York 14627 
}
 \affiliation{School of Physics and Astronomy, Rochester Institute of Technology, Rochester, New York 14623}

\author{Molly Griston}
\affiliation{
 Department of Physics and Astronomy, University of Rochester, Rochester, New York 14627 
}

\author{Jes\'{u}s Botello}
\affiliation{Department of Physics, The University of Texas at Austin, Austin, Texas 78712} 

\author{Benjamin Zwickl}
 \affiliation{School of Physics and Astronomy, Rochester Institute of Technology, Rochester, New York 14623}

\date{\today}

\begin{abstract}


Understanding how physicists solve problems can guide the development of methods that help students learn and improve at solving complex problems.  Leveraging the framework of cognitive task analysis, we conducted semi-structured interviews with theoretical physicists ($N=11$) to gain insight into the cognitive processes and skills that they use in their professional research.  Among numerous activities that theorists described, here we elucidate two activities that theorists commonly characterized as being integral to their work: making assumptions and using analogies.  Theorists described making assumptions throughout their research process, especially while setting their project's direction and goals, establishing their model's interaction with mathematics, and revising their model while troubleshooting.  They described how assumptions about their model informed their mathematical decision making, as well as instances where mathematical steps fed back into their model's applicability.  We found that theorists used analogies to generate new project ideas as well as overcome conceptual challenges.  Theorists deliberately sought out or constructed analogies, indicating this is a skill students can practice.  When mapping knowledge from one system to another, theorists used systems that shared a high degree of mathematical similarity; however, these systems did not always share similar surface features. We conclude by discussing connections between the ways theorists use assumption and analogy and offering potential new avenues of research regarding applications to instruction.


\end{abstract}

\maketitle

\section{\label{sec:Introduction}Introduction}

Teaching students to become skillful problem solvers is one of the primary responsibilities of physics departments and has been the subject of research in physics education for decades \cite{reif1976teaching, reif1982knowledge, docktor2014synthesis, hsu2004resource}.  Developing expert-like problem solving skills is especially important for undergraduate students with aspirations of becoming STEM professionals, since those are the skills that they will utilize on a daily basis in their future careers \cite{american_institute_of_physics_employment_2020}.  To meet this crucial goal, tremendous progress has been made generating physics education frameworks to help students become better problem solvers \cite{heller1984prescribing, van1991learning, heller1992teaching1, huffman1997effect}.  Numerous problem types have been designed to promote physical reasoning in diverse situations, including context-rich problems \cite{heller1992teaching}, synthesis problems \cite{ding2011exploring}, and categorization tasks \cite{dufresne1992constraining, hardiman1989relation}, among others.  

The effort to formulate such a wide range of problem types for students stems from the fact that problem-solving strategies can be markedly different depending on context.  Although standard textbook problems often require substantial physics knowledge to solve correctly, they are disparate from the types of problems with which researchers grapple on a daily basis.  Researchers typically work on open-ended and ill-defined problems, while the exercises that students find in their textbooks often have clear answers and a relatively straightforward path to a solution.  Thus, students who receive only limited opportunities to develop their problem-solving skills beyond those used in well-defined exercises may have difficulty solving more complex real-world problems.  


Relying solely on traditional pedagogical practices and instructional problems may hinder physics students' acquisition of important skills by rendering key aspects of expert practice ``invisible" \cite{schoenfeld1985mathematical, collins1987cognitive}.  Little of what experts do is obvious or easily copied by non-experts.  Students watching their instructors solve textbook problems are unlikely to pick up on many of their subtle processes, just as a novice golfer is unlikely to significantly improve their swing simply by watching Tiger Woods \cite{adams2015analyzing}.  Hence, ample research on modeling instruction \cite{hestenes1987toward, wells1995modeling, brewe2008modeling} and labs \cite{etkina2010design, zwickl2013process} have emphasized the importance of engaging students in practices that mirror those performed routinely by expert physicists.  Still, despite the known importance of engaging students in good problem-solving techniques, there have been few efforts to methodically characterize the complex problem-solving processes employed by physicists in their authentic research.  Rather, many physics education research studies on expert-like problem-solving focus on how experts (typically faculty and graduate students) solve well-defined, introductory-level physics problems \cite{docktor2014synthesis}.

Research specifically examining how physicists solve real-world problems related to their research is comparatively limited \cite{park2009analysis, leak2017examining, hu2019characterizing, price2021detailed, jonassen2006everyday, griston2021light}.  Prior studies have provided little information about the context in which the physicists were performing problem-solving activities, yielding limited insight into how and why physicists made the decisions that they made.  Furthermore, studies have not differentiated between strategies used by experimentalists and theorists, leaving ambiguity as to which skills were utilized more often in lab settings or while solving theory-driven problems.  While the skills utilized by theorists and experimentalists are likely similar, they are certainly not identical; theorists seldom have to fix a broken apparatus.  \citeauthor{hestenes1992modeling} differentiates between the types of activities undertaken by theorists and experimentalists in his research on modeling instruction, positing that ``The theorist builds the conceptual world (a world of the possible), while the experimentalist explores the physical world (the world of the actual)" \cite{hestenes1992modeling}.  This distinction has important implications for physics education, since much of the content in traditional undergraduate physics courses is centered on theory and does not have an experimental component. 

The current study aims to fill gaps in the current literature by examining theoretical physicists' problem solving in the context of a single real-world research project.  We conducted semi-structured interviews with $N=11$ self-identified theoretical physicists using an interview protocol inspired by the Critical Decision Method of cognitive task analysis and Applied Cognitive Task Analysis \cite{hoffman1998use, militello1998applied, crandall2006working}.  Interviews involved multiple-pass retrospection on a theorist's recently completed project guided by probe questions designed to elicit expert recollection of both what happened during an event and why it occurred.

We began coding the interview transcripts with the broad goal of outlining the steps that theorists take to solve problems in their research.  We also generated process diagrams (see Figure \ref{fig:callout}) to visually represent the theorists' processes and aid in identifying important themes in the data.  However, after several rounds of coding and analysis we observed several important processes that warranted individual attention.  Specifically, across all interviews, making effective assumptions and using analogies in the problem solving process were clearly important.  These activities are the subject of this paper.

Once we identified assumptions and analogies as important aspects of the authentic scientific inquiry process employed by theorists, we refocused our analysis to further investigate these topics.  In particular, we sought to answer the following research questions:
    \begin{enumerate}
        \item What kinds of assumptions do theorists make?
        \item What roles do assumptions play in theorists' problem solving processes?
        \item Why do theorists use analogies in their work?
        \item How do theorists identify analogies?
    \end{enumerate}

By answering these questions, we hope to make several ``invisible" expert processes visible to students and faculty alike.  Elucidating aspects of theorists' problem-solving practices that instructors can integrate into undergraduate and graduate curricula will allow teachers to better leverage expert approaches for supporting student learning.  We conclude by offering suggestions for avenues of research to encourage students to engage in more expert-like ways with assumptions and analogies.

\section{\label{sec:Background} Background}

\subsection{Expert problem solving}



Many physics education research studies on expert-like problem-solving practices focus on how experts (faculty and graduate students) solve introductory-level problems \cite{docktor2014synthesis}.  This research often contrasts the ways that expert and novice physicists solve problems, and has found distinct differences between the knowledge structures and procedures that undergraduate students and faculty use while problem solving \cite{larkin1980expert, chi1981categorization, dufresne1992constraining, bransford2000people, clement1988observed}. 

Although comparing expert and novice problem-solving practices has provided many important revelations with regard to teaching and learning physics, there are several notable limitations to this type of work.  For instance, these studies are typically focused on solving well-defined introductory problems, which are not representative of the kinds of problems physicists solve in their professional research.  Furthermore, some research suggests that the expert-novice differences can become less noticeable under certain circumstances, such as when experts are given more difficult problems \cite{hsu2004resource} or when novices undertake more open-ended, context-rich problems \cite{ogilvie2009changes}.  \citeauthor{ho2001some} is a relevant exception in the expert-novice literature and explicitly studied differences in solving ill-defined problems, finding that experts use constraining assumptions in order to break down complex problems into simpler ones \cite{ho2001some}.

Work examining problem-solving processes of physicists in real-world scientific contexts is comparatively limited.  In one expert self-reflection, \citeauthor{wieman2015comparative} offered a list of mental tasks specifically associated with experimental physicists carrying out tabletop research \cite{wieman2015comparative}.  In another study focusing specifically on physicists,  \citeauthor{park2009analysis} performed retrospective interviews with three experimentalists and three theorists, asking participants to recall why and how they worked on various projects.  The authors identified five primary processes that the physicists reported using in their authentic research: defining and preparing research problems, generating hypotheses, designing research, executing research, and drawing conclusions \cite{park2009analysis}.  Among a variety of sub-processes that the authors identified, they noted that testing assumptions was a crucial component of executing research.  In another study focusing on physics graduate students, \citeauthor{leak2017examining} characterized the types of problems that students face in their PhD work, as well as some of the strategies students used to solve different types of problems.  Strategies included getting help from professors and peers, using estimation, applying test cases, and using trial and error, among many others \cite{leak2017examining}.  

Other investigations of expert problem solving in realistic science environments have either focused on specific disciplines outside of physics or on STEM professionals as a whole, and have employed a variety of data collection methods.  These include diverse studies ranging from expert self-reflection on mathematical problem-solving \cite{polya1945how} to interviewing experts in engineering workplaces \cite{jonassen2006everyday}.  In this analysis we used similar methods to \citeauthor{price2021detailed}, who leveraged cognitive task analysis approaches to conduct retrospective interviews with 52 STEM professionals about how they solved a particular problem in their research (see Section \ref{sec:Method} for details on cognitive task analysis methods).  The authors identified a set of 29 primary decision points that experts use across disciplines, such as deciding what problems are important in their field and which predictive framework to use \cite{price2021detailed}.  

A notable exception to studies using interviews as the method of data collection is \citeauthor{dunbar1997scientists}, who used direct observation of biologists in their labs \cite{dunbar1997scientists}.  One of the major findings in this study was the importance of analogical reasoning in the expert problem-solving process.  The study found that all of the observed biology researchers used analogies regularly in their work, and that they were often employed in an attempt to explain an observed phenomenon \cite{dunbar1997scientists}.  A subsequent meta-analysis of expert problem solving also listed analogy usage as an important tool for researchers \cite{chinn2002epistemologically}.

\subsection{Assumptions}

\subparagraph{Defining assumptions} Several publications in science education literature provide definitions for ``assumptions."  For instance, \citeauthor{scates1935types} defines an assumption as ``a mental datum which is not fully established, but which is used as a basis for continuing the thought or study" and has the property that it ``covers all facts, principles, or other concepts, the truth of which is taken for granted for particular purposes without insistence of specific proof" \cite{scates1935types}.  In a study on the use of assumptions in solving real-world problems,  \citeauthor{fortus2009importance} also adopts this definition, but elaborates on their purpose in solving ill-defined problems.  Based on prior work \cite{reitman1964heuristic},  \citeauthor{fortus2009importance} differentiates ``well-defined" and ``ill-defined" problems by the degree to which the problems are constrained.  He argues that making assumptions serve to constrain an ill-defined problem and ``make it ``more" defined, limit the size of its solution space, turn it into a well-defined problem" \cite{fortus2009importance}. 

Thus, in the context of scientific research, assumptions are defined by two main features.  One is their utility to the researcher in continuing the problem-solving process by making the problem tractable, and the other is their relation to things that the researcher believes to be ``true" in a problem.  However, we believe the usage of phrases such as ``constrain" imply that assumptions are always used to simplify a problem, even though we can easily conceive of instances in which a researcher might want to break an assumption in order to make a model more complex. In order to avoid presupposing the roles that assumptions play in theorists' problem-solving processes, in this paper we focus primarily on the second aspect of the definition that regards assumptions as features of a model that researchers take for granted to be true.

\subparagraph{Role in modeling} The significant role that making effective assumptions plays in physics problem solving is clearly emphasized by education research on modeling instruction.  Of the scientific practices, modeling has received a great deal of study and attention within science and physics education because it involves a process that seems to encapsulate much of what it means to ``do science."  Entire curricula have been designed that center on modeling, and assumptions are always an integral part of the modeling process \cite{etkina2006scientific, wells1995modeling, brewe2008modeling}.  In one modeling framework designed specifically for experimental physics, making ``assumptions and simplifications to make the model tractable" is an explicit step situated in the overall framework \cite{dounas2018modelling}.  Although the importance of having students highlight assumptions in their models is known, further work would help to provide more clarity as to when and why particular assumptions are made.

Much work in physics education research on modeling has built on the foundation of \citeauthor{hestenes1987toward}, who argues that having students recognize and make assumptions mimics authentic problem solving practice better than traditional instruction \cite{hestenes1987toward, hestenes1992modeling, hestenes1993modeling, halloun1987modeling, halloun1996schematic}. \citeauthor{hestenes1987toward} notes that ``textbooks and instructors have been known to write down equations of motion without helping students identify all the assumptions involved," while modeling instruction helps to make assumptions explicit \cite{hestenes1987toward}.  \citeauthor{hestenes1993modeling} defines a model as consisting of four primary features: constituents, descriptors, laws, and interpretation \cite{hestenes1993modeling}.  Constituents are the things of interest in the model as well as the things in their environment.  Descriptors are the attributes of the constituents and represent properties of the things in question.  Our interpretation relates the descriptors in the model to the properties they represent in the world.  These descriptors can be fixed (e.g., a projectile's mass) or can change (e.g., a projectile's velocity), and interactions between two things in the model change their properties according to laws (e.g., forces defined by Newton's Second Law).  We use a variation of this model description later in order to classify types of assumptions that theorists make (results of this classification are presented in Section \ref{sec:assumptionkinds}).

\subsection{Analogy}

\begin{figure}
\includegraphics[]{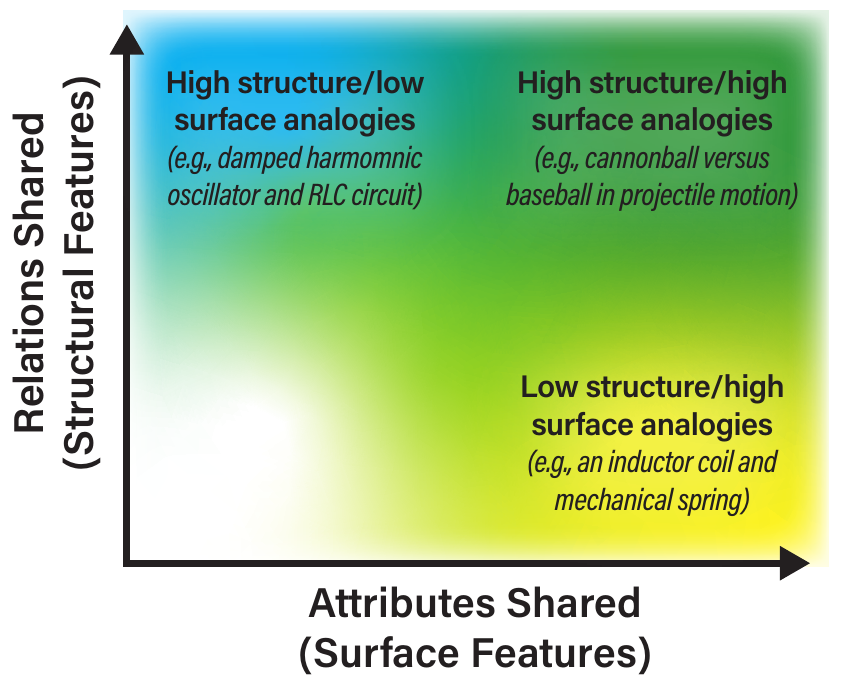}
\caption{\label{fig:simspace} A visual representation of the space of analogy types that we use to classify theorists' analogies, based on the extent to which structural (vertical axis) and surface (horizontal axis) are mapped from base to target.  Colors demonstrate that this space is a spectrum rather than a set of distinct categories.  Concept for figure adapted from \citeauthor{gentner1997structure} (1997) \cite{gentner1997structure}.}
\end{figure}

The power of analogy in solving problems is well-known to practicing physicists, and has been acknowledged by historical figures such as Maxwell and Oppenheimer \cite{maxwell_2011, oppenheimer1956analogy}.  In this section, we provide an overview of existing theory regarding the definition of analogy and the cognitive processes involved in invoking analogous reasoning.

Across a variety of contexts and disciplines, authors broadly agree that an ``analogy" consists of mapping some knowledge from a base to a target \cite{gentner1983structure, holyoak1989analogical, brown1989overcoming, dunbar1997scientists, podolefsky2006use}.  The target is the problem that the scientist is trying to explain, while the base is another piece of knowledge that the scientist is using to understand the target.  As an example, when mapping features of the solar system (the base) to the Rutherford atom (the target), one might map the relations between the sun, planets, and gravitational force in the solar system to the relations between the nucleus, electrons, and the Coulombic force in the atom.  Notably, despite these structural similarities, there is no physical resemblance between an electron and a planet (i.e., one is large and visible, the other is small and invisible).  These systems do not share any surface attributes; however, it is possible to make a comparison predominantly based on surface-level features rather than structural ones. For example, one might compare the attributes of an inductor coil to a mechanical spring (i.e., both  metallic, wiry, and curled in a helix), while ignoring their purpose in physical systems and their related mathematics.  

These examples demonstrate that the features of a problem that are mapped from the base to the target occur along two axes: \textit{relations shared} and \textit{attributes shared} (depicted graphically in Figure \ref{fig:simspace}) \cite{gentner1997structure}.  The ``relations shared" axis refers to common structural level features between the base and target, whereas the ``attributes shared" axis refers to surface level features. As the base and target share more of both relational (structural) and attribute (surface) features, they converge into the exact same system.  For instance, a student might posit that a cannonball in projectile motion will have the same flight as a baseball in projectile motion since the two systems both appear visually similar (i.e., a round object being launched into the air) and the same forces and laws describe their motion.  The bifurcation of feature types into structural and surface characteristics is outlined in cognitive psychologist Dedre Gentner's theory of ``structure mapping" \cite{gentner1983structure}, a foundational work on the theory of analogy. 

Some authors use more stringent criteria than others to define which types of knowledge mappings constitute ``analogies" and which ones do not, and these definitions can vary significantly among researchers \cite{brown1989overcoming,podolefsky2006use,dunbar1997scientists,taber2001analogy}. Some physics education research literature does not discern between the different types of features mapped between systems at all, instead opting to define analogy simply as \textit{any} mapping of similar features from base to target \cite{podolefsky2006use, korhasan2019should}.  However, this approach has the obvious disadvantage of failing to illuminate many of the nuanced ways that individuals make and use analogies, as well as the ways that usage differs between students and experts.  

On the other hand, Gentner distinguishes analogy (sometimes called ``relational metaphors") from other types of comparisons by requiring analogies to share many structural features but few surface features (upper-left quadrant of Figure \ref{fig:simspace}) \cite{gentner1983structure}.  Under Gentner's definition, a comparison between a damped harmonic oscillator and an RLC circuit would constitute an analogy, but a comparison between a cannonball and baseball in projectile motion would not since they also share many surface features.  Although the same underlying cognitive process is used (i.e., mapping features from base to target), Gentner only uses ``analogy" to describe a particular subset of comparisons.  

Still others have adopted theoretical frameworks that retain the two axes of mapped features, but define all types of comparisons as ``analogies" \cite{harrison1993teaching, gokhan2012effects,dilber2008effectiveness,dunbar1997scientists}.  For instance, \citeauthor{gokhan2012effects} write, ``Many students do not realize that analogies operate on two levels.  In simple appearance matches or descriptive analogies, one or more superficial attributes of the analog corresponds with the target, whereas true inductive analogies share both superficial and higher order causative relations" \cite{gokhan2012effects}.  Thus, while differentiating between the types of knowledge mapped from base to target, the authors maintain that the comparison is still an analogy.  This is broadly in alignment with \citeauthor{dunbar1997scientists}, who classified all comparisons that biology researchers made as analogies but differentiated whether analogies were ``non-biological," ``other organism," or ``within organism"  \cite{dunbar1997scientists}.  Relating these categories to Figure \ref{fig:simspace}, ``non-biological" and ``other organism" examples likely occur in the upper-left (high structure/low surface) quadrant, while ``within organism" examples align more closely with the upper-right quadrant (high structure/high surface).

In this work, we sought to use a theoretical lens that captures the nuance of the types of knowledge mapped and facilitates a useful classification scheme for theorists' analogies.  Therefore, taking into account the diverse ways that analogy is treated in the literature, we adopt the framework presented in Figure \ref{fig:simspace}.  This framework considers all kinds of comparisons to be a type of analogy, and we categorize them based on the features being mapped (high structure/high surface, high structure/low surface, low structure/high surface, with the acknowledgement that each of these regimes exist on a spectrum).  

We believe that presenting each different case of knowledge mapping as an analogy has the benefit of eliminating potentially distracting debate over what to label a particular comparison and redirecting that focus onto the actual knowledge being mapped.  Rather than grappling with the question ``Is this comparison an analogy or not?", this approach forefronts the particular features of a problem that theorists are mapping onto another, and encourages consideration of why certain knowledge mapping is useful in different situations but not others.

\section{\label{sec:Method}Method}

\subsection{Cognitive Task Analysis}
We used several methods outlined in the Cognitive Task Analysis (CTA) literature to identify aspects of theorists' research processes that account for their skilled performance in real-world settings.  CTA refers to the broad variety of methods used to collect data on expert knowledge, analyze that data, and subsequently represent the expert knowledge in a useful way \cite{crandall2006working}.  The goal of a CTA study is to systematically characterize the decisions made by experts in their authentic working environments in such a way that is amenable to transferring that knowledge to others.  

The primary differences between CTA methods stem from the ways that these steps are executed.  For example, knowledge may be elicited from experts via interviews, self-reports, or direct observations, with each method offering certain advantages and disadvantages.  In particular, interviews offer an efficient way of gathering data on long-term research projects, especially when compared to a resource-intensive method like direct observation.  However, they may not offer the same level of richness that well-designed observational studies provide, and interviewees may not be able to fully and accurately recall (or may not be aware of) important processes in their research.  Other decisions regarding knowledge elicitation revolve around whether to collect data on tasks that the subjects performed in the past or are presently undertaking, whether those tasks are real or hypothetical, and whether the tasks are representative of typical day-to-day activities or rare/challenging activities \cite{crandall2006working, clark2008early}.  

CTA methods also differ in the ways that the elicited knowledge is analyzed and represented.  Interview data might be analyzed by cataloguing cues and patterns that contribute to expert performance or by identifying broad themes within the data. Representations may include textual descriptions, tables, or charts and graphs.  Due to the vast number of methods available to researchers seeking to conduct a CTA, practitioners often combine and adapt a range of methods to explore the particular cognitive phenomenon that they wish to better understand \cite{crandall2006working}.

In this study, we opted to elicit expert knowledge from theoretical physicists through semi-structured interviews.  Since we sought to characterize aspects of theorists' problem-solving throughout the course of an entire research project, interview prompts asked experts to give a retrospective account of a single research project that had recently been completed.  Interviews allowed us to efficiently gather data on a theorist's project even if it spanned many months or years.  We designed an interview protocol based largely on the Critical Decision Method of task analysis \cite{hoffman1998use} and Applied Cognitive Task Analysis \cite{militello1998applied}, which are both methods of CTA constructed to capture real-world decision making processes of experts on a specific task.  Examples include how doctors and nurses make decisions regarding patient care (critical decision method) \cite{gazarian2010nurse, fackler2009critical} and how pilots make certain decisions while flying (applied cognitive task analysis) \cite{seamster2017applied}.  
\begin{figure*}[t]
\centering
\includegraphics[]{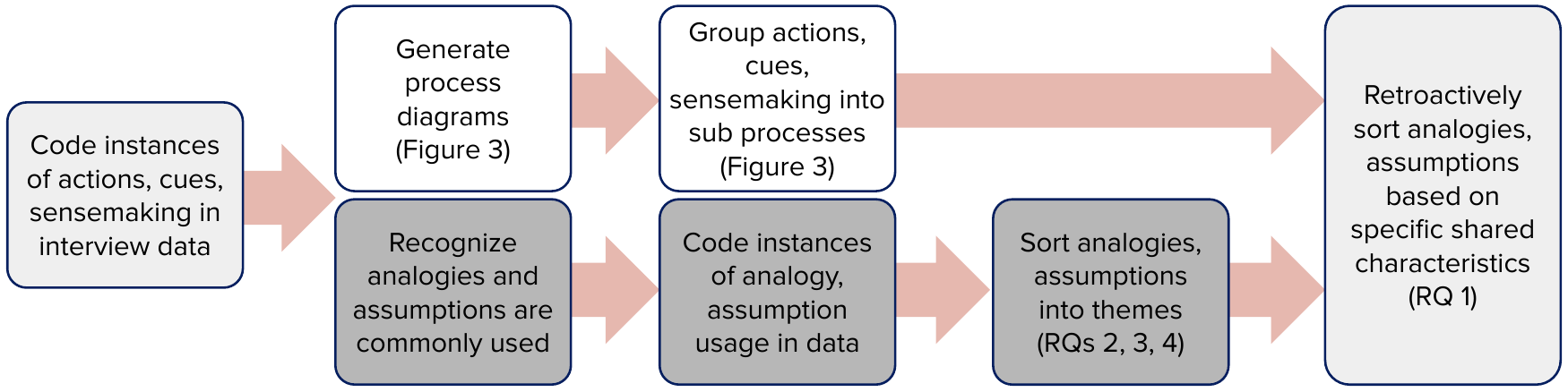}
\caption{\label{fig:timeline} A flowchart summarizing the rounds of analysis performed on the interview data.}
\end{figure*}
Theorists gave accounts of their project several times, progressively deepening in detail so that we could gain a comprehensive and contextually rich account of the project.  To prompt a first pass recounting of a recently completed research project, we asked the interviewee to describe what they recalled as being the main stages of the process from start to finish (``Think about the process you took to complete this project, start to finish. Can you split this into ten stages or so? These don't have to be detailed steps, just a general process.").  As the interviewee described the main stages of their project, we transcribed important steps in a shared document called a ``Task Diagram," which was viewable by both the interviewer and the interviewee.  We then prompted the interviewee to go back through the main tasks enumerated in the Task Diagram and to describe them in more depth.  In this stage of knowledge elicitation, we also asked a series of in-depth probe questions designed to capture the cognitive processes used by the theorists at various stages of the project listed in the Task Diagram (e.g., ``How did you know this project was doable and significant?").  Probe questions included inquiries specifically related to math (e.g., ``When you said you used math [in this step], in what form or representation was it used and why did you choose that representation?"), as well as non-mathematical inquiries (e.g., if the theorist collaborated on their project, ``How do you and the collaborator benefit from the collaboration?").






The full interview protocol is included in the supplemental material \cite{supplementalMats}. Interviews typically lasted 1 to 1.5 hours. Each interview was recorded and transcribed using Zoom. After each of the interviews, we corrected the automated transcription for errors and punctuation to generate accurate transcripts.  These transcripts became the subject of our analysis.

\subsection{Coding and Analysis}

Transcripts underwent several rounds of coding and analysis as we progressively sharpened the focus of the study (see Figure \ref{fig:timeline} for a summary of our analysis steps). We first began coding interviews using Dedoose software \cite{dedoose} with the broad goal of identifying important aspects of theorists' work as well as how they make important decisions.  Thus, in the first phase of analysis we coded interviews using a set of three process codes to characterize theorists' problem solving processes: actions, cues, and sensemaking (see Table \ref{tab:actioncodes}).  Actions described the particular decisions and performative aspects of the research process, while sensemaking described aspects of the theorists' cognitive processes \cite{odden2019defining}.  Sensemaking often involved phrases like ``thinking," ``realizing," ``knowing" and ``discovering" and described processes that occurred in the theorist's mind.  Cues referred to any aspects of the problem or thought processes that led to subsequent actions or instances of sensemaking.  In some cases, outcomes of one action became cues of the theorist's next action.

By using cue as a sub-code of actions and sensemaking, we were able to link cues to the specific action or cognitive process that they elicited.  Once a subset of five of the 11 interviews were coded using this framework, we exported our codes and sub-codes as a spreadsheet that could be manipulated using the R programming language \cite{baseR_package}.  We wrote an R program to generate tables organizing instances of actions and sensemaking alongside their accompanying cues for each individual interview.  Printing paper copies of these allowed us to begin organizing the data into process diagrams that visually displayed the approximate temporal order of the action/cue/sensemaking codes that we identified for each individual theorist.  From these, we generated refined process diagrams using the software Lucidchart \cite{lucidchart} for each theorist (see Figure \ref{fig:callout}).  

We first analyzed the process diagrams by grouping parts of each theorist's diagram into sub-processes that we identified (e.g., an idea generation, preliminary design and analysis, etc.).  We defined these based on the common goals of each phase.  As an example, we identified common steps that theorists took to determine whether a new project idea was doable or not, and classified them as being part of a preliminary design and analysis sub-process.  This part of our analysis helped to advance our broad goal of characterizing the types of activities that theorists undertook.  It also allowed us to gain a better understanding of the data and to begin referring to approximate phases of theorists' overall process (results of this analysis are presented in Section \ref{sec:initialcodes}).  

While analyzing the process diagrams in this way, we began to observe a number of themes related to similar instances of actions/cues/sensemaking.  For instance, we quickly noted similarities in the cues that influenced a theorist's willingness to go through with a project (e.g., confident in ability, students were excited about the idea, is doable with long-distance collaborator). 

\begin{table}[t]
\centering
\def\arraystretch{1.5}
\begin{tabularx}{\columnwidth}{ZZ}
\hline
\multicolumn{1}{l}{\textbf{Process Codes}} & \multicolumn{1}{c}{\textbf{Description}} \\ 
\hline \hline
\textit{Action} & 
Performative aspects of the research process, such as making specific decisions and executing physical tasks \\

\textit{Cue} & 
Prompts that lead theorists to make certain decisions or think a particular way about the problem  \\

\textit{Sensemaking} & 
Describes aspects of the cognitive process undertaken by theorists, often denoted by phrases like ``thinking," ``realizing," or ``knowing"  \\ 

\hline \hline
\end{tabularx}
\caption{\label{tab:actioncodes} Short descriptions of the process codes that were used during the first phase of analysis.}
\end{table}

Two of the observed themes were theorists' frequent usage of assumptions and analogies in the scientific inquiry process, which became the focus of our analysis.  Thus, rather than deciding to more comprehensively characterize all aspects of theorists' problem solving processes, we shifted goals and opted to deeply examine theorists' use of analogies and assumptions.  We therefore generated a new set of research questions (presented in the introduction) and began a revised coding process focused specifically on these questions.  Results related to assumptions are discussed in detail in Section \ref{sec:Assumptions} and analogies are discussed in Section \ref{sec:Analogy}.

From the broader set of excerpts that were coded with the action/cue/sensemaking framework, we used a revised coding scheme to isolate instances of assumption and analogy usage within these excerpts.  We leveraged the definitions of assumptions and analogies outlined in Section \ref{sec:Background} to define the assumption and analogy codes in our revised codebook.  Several themes emerged from this process and we began to progressively refine them with subcategories as all 11 interviews were coded (see Tables  \ref{tab:assumptionroles} and \ref{tab:analogyroles} in the Appendix for lists of these codes).

\begin{figure*}[t]
\includegraphics[]{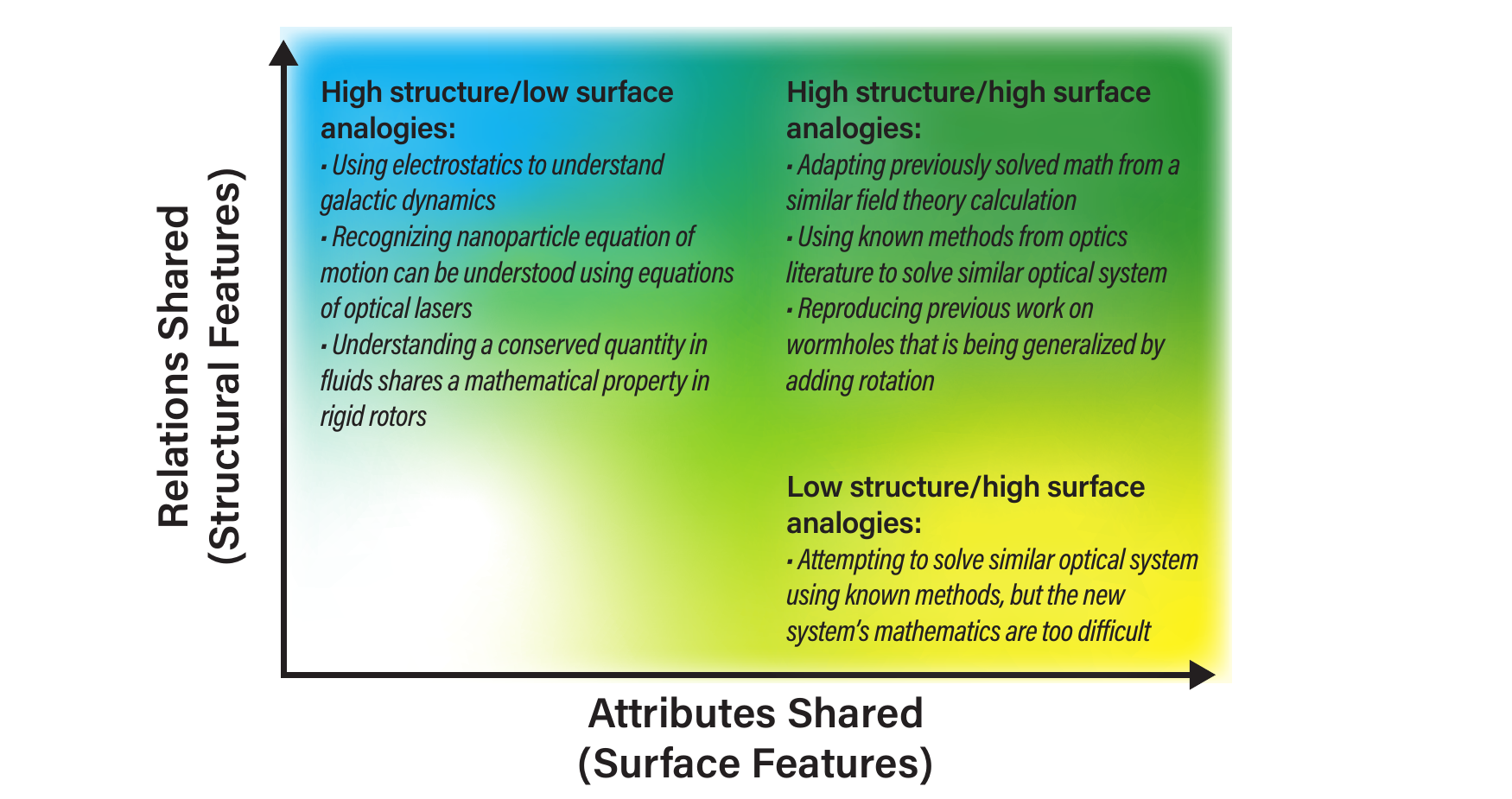}
\caption{\label{fig:analogyexamples} Several instances of analogy usage that we identified, mapped onto the analogy space presented in Figure \ref{fig:simspace}.  Analogies in the upper left corner labeled high structure/low surface were mathematical analogies between systems from ``distant" domains of physics (e.g., electrostatics and galactic dynamics).  The upper right side of the diagram illustrates analogies between highly similar systems on both levels (e.g., adapting previously solved calculations from within the theorist's subfield).  Other analogies fell on a spectrum between the upper right and lower right corner, depending on the degree to which the theorist needed to change their mathematical approach to the new problem.}
\end{figure*}

Since the first themes we identified were strongly linked to our first pass action/cue/sensemaking coding scheme, these themes primarily corresponded to research questions 2, 3 and 4 regarding why and how theorists actively use and identify assumptions and analogies.  However, while working with the data to answer these questions we began to recognize other patterns in the \textit{types} of assumptions and analogies theorists made.  Thus, we retroactively undertook a new round of coding to sort assumptions and analogies into categories based on certain shared characteristics.  We used a variation of \citeauthor{hestenes1993modeling}' description of conceptual models in science (outlined in Section \ref{sec:Background}) to classify assumptions (results presented in Table \ref{tab:assumptiontypes}) \cite{hestenes1993modeling}.  Similarly, we used the analogy framework shown in Figure \ref{fig:simspace} to classify the types of analogies theorists make (results presented in Figure \ref{fig:analogyexamples}).  Although these analyses occurred last chronologically during our analysis, we present them first in the Results section to give readers a basic framework for thinking about the types of assumptions theorists discussed during the interviews.

Throughout the coding process we used several approaches to produce a valid and reliable coding scheme.  One researcher (M.V.) was primarily responsible for implementing the first pass action/cue/sensemaking coding scheme, as well as the creation of the subsequent code book used to identify analogies and assumptions.  However, identification of these themes and refinement of the codes occurred during weekly meetings with another member of the research team (B.Z.), as well as through discussions with others outside the research team.  At several points during the coding process, a group of discipline-based education researchers external to the research project were asked to code random excerpts from the interviews using the most updated code book. Disagreements in coding were resolved and discussions informed numerous refinements to the code book.

\subsection{Interview Demographics}
\begin{table*}[t]
\centering
\setlength{\tabcolsep}{0.25em}
\def\arraystretch{1.2}
\begin{tabular}{P{0.15\textwidth}P{0.35\textwidth}P{0.5\textwidth}}
\hline
\multicolumn{1}{l}{\textbf{Theorist}} & \multicolumn{1}{c}{\textbf{Subfield}}                & \multicolumn{1}{c}{\textbf{Project Description}}   \\ \hline \hline
Dr. Agarwal & Cosmology, Fundamental Quantum & Determining whether matter distributions generating gravitational fields with nonzero curl affect predictions of an alternative model to Newtonian gravity
\\
Dr. Bahl  & String Theory & Investigating effects of rotation on various characteristics of wormholes
\\
Dr. Costa  & Biophysics & Using analytical and simulation approaches to understand implications of multiple filament structure growth in cells                              
\\
Dr. Dunn & Astrophysics & Understanding the origins of magnetic fields in large scale astrophysical objects       \\
Dr. Erdogan & Particle Physics, Quantum Field Theory & Discovering new features of classical fluid mechanics via an analogy with the rigid rotor        \\
Dr. Fisher & Quantum Optics & Establishing new structures that act like fundamental circuit elements for larger networks                                                
\\
Dr. Garcia & Quantum Optics & Constructing a new laser based on mechanical vibrations via analogy with standard optical laser physics                                      \\
Dr. Huang & Computational Astrophysics & Describing  the collisions to supermassive black holes and calculating of the dynamics of accretion disks                            \\
Dr. Irvine & Biophysics & Modeling mitochondrial population dynamics to predict conditions leading to healthy or unhealthy outcomes                                         \\
Dr. Jackson & Particle Physics, Phenomenology & Defining experimental signatures that would establish evidence for a new particle generated by a particular mechanism                        \\
Dr. Khan & Quantum Optics & Constructing a new laser based on mechanical vibrations via analogy with standard optical laser physics                                              \\ \hline \hline
\end{tabular}
\caption{\label{tab:demographics} Table of theorists' subfield and a description of the research projects they described in their interviews.  Of the $N=11$ total participants, $N=6$ verbally described their gender as ``male," $N=2$ ``female," and $N=3$ as a ``woman."  The institutions that the theorists represented included a small undergraduate liberal-arts college, two large public 4-year universities, and two large private 4-year universities.}
\end{table*}

We interviewed $N=11$ theoretical physics faculty members from five colleges and universities.  Although physicists are often divided into experimentalists and theorists, we acknowledge that there is no strict dividing line between these groups.  To avoid making judgments on who qualifies as a ``theorist," our selection criteria for interview subjects was that they self-identified as a theoretical physicist.  As a result, interviewees differed with regard to their use of computation and overlap with experimental work.  While some theorists described themselves as largely analytical, only using tools such as Mathematica to quickly solve certain equations or check their pencil-and-paper results, others relied heavily on large-scale simulations in their research.  Some utilized a combination of analytical and numerical approaches.  Several interviewees worked on abstract mathematical problems with few experimental implications, while others worked on projects tightly coupled to the work of experimental groups.  Table \ref{tab:demographics} provides brief descriptions of each theorist's self-identified subfield and a short description of the project they detailed in their interview. 

One implicit limitation of all studies on expert-like behavior is that they rely on the assumption that the present standard of how practicing physicists operate is how they \textit{should} operate.  Making this limitation explicit allows readers to critically evaluate the practices uncovered in these studies, particularly with regard to their implications for issues of representation in physics.  Physics has historically underrepresented women and BIPOC students and faculty \cite{traxler2016enriching}, so failing to take into account the experiences from a diverse set of participants in expert studies risks creating a false standard of ``expertise" based on a limited subset of the physics community.  Mindful of this limitation, we attempted to recruit theorists from a variety of backgrounds and identities.  

We began recruiting participants via email at our home physics departments, then added several more participants through suggestions from the initial set of interviewees.  In order to honor and acknowledge each researcher's experience and background, we asked at the beginning of each interview if there were any identities, including age, gender and ethnicity, that they wished to share.  Any details shared were voluntarily disclosed as answers to a prompt in the interview protocol (``Could I ask you about your age, gender, ethnicity, and/or any other identities you'd like to share?").  Of the $N=11$ total participants, $N=6$ verbally described their gender as ``male," $N=2$ ``female," and $N=3$ as a ``woman."  Many identified the specific country they were born; these included India ($N=4$), Peru ($N=1$), Italy ($N=1$), China ($N=1$) and Turkey ($N=1$).  In total, $N=4$ identified as ``White."  Subjects were at varying stages of their careers at the time of the interviews.  One had been hired within a year of the interview taking place while another was finishing a post-doctoral appointment and had accepted a faculty position.  Others were tenured with large and active research groups, while one was tenured but no longer had an active research group.  Researchers worked at several institution types, including a small undergraduate liberal-arts college, two large public 4-year universities, and two large private 4-year universities.

\section{\label{sec:Results}Results}
Results begin with Section \ref{sec:initialcodes}, which describes the sub-processes that we identified within theorists' overarching research projects.  We move on to our analysis of assumptions in Section \ref{sec:Assumptions}.  This section is split into two broad categories, corresponding to our research questions.  Section \ref{sec:assumptionkinds} is dedicated to RQ1 and discusses the kinds of assumptions that theorists make as well as when those assumptions occur in the overall problem solving process.  Then in Section \ref{sec:assumptionroles}, we address RQ2 by identifying the roles that assumptions played in theorists' problem solving processes.  Section \ref{sec:Analogy} is dedicated to analysis of theorists' use of analogies in their research, where we begin by outlining the types of analogical reasoning we observed in our data by mapping instances of analogy use onto the analogy space presented in Figure \ref{fig:simspace}.  This is followed by Section \ref{sec:analogywhy}, which addresses RQ3  regarding why theorists use analogies.  Lastly, Section \ref{sec:analogyhow} is devoted to RQ4 and describes how theorists find analogies.

\subsection{\label{sec:initialcodes}Sub-processes of theorists' problem solving}

\begin{figure*}
\centering
\includegraphics[]{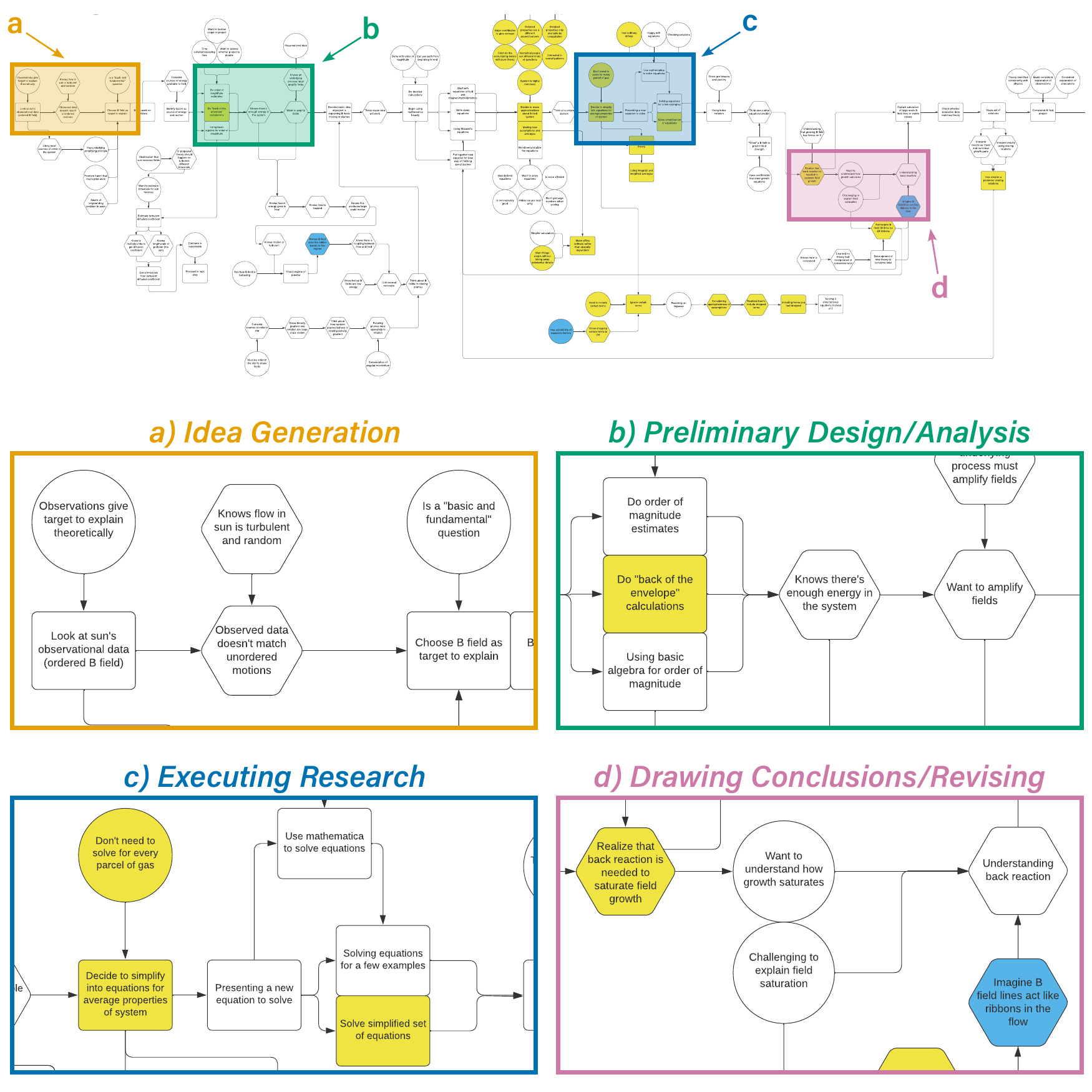}
\caption{\label{fig:callout} A sample process diagram that was generated using the actions, cues, and sensemaking process codes, with callouts indicating representative examples of the different sub-processes.  Rectangles represented actions, circles represented cues, and hexagons represented instances of sensemaking.  Shapes shaded yellow indicate that they are related to assumptions while blue indicates a relation to analogy.  Visually, arrows connecting the action/cue/sensemaking bubbles point in all directions on the diagram, indicating that theorists frequently jump between different points in their overall process.  In this specific diagram, the prevalence of yellow shapes illustrates that this theorist frequently described aspects of their process related to making effective assumptions.  Figure generated with Lucidchart \cite{lucidchart}.}
\end{figure*}

\begin{table}[b]
\centering
\def\arraystretch{1.5}
\begin{tabularx}{\columnwidth}{ZZ}
\hline
\multicolumn{1}{l}{\textbf{Problem solving sub-process}} & \multicolumn{1}{c}{\textbf{Description}} \\ \hline \hline

\textit{Idea Generation} & 
Recognizing a potential new avenue of research, defining a new research question  \\

\textit{Preliminary design/analysis} & 
Planning the research, gathering collaborators, conjecturing about possible solutions, deciding whether idea is doable, order of magnitude calculations \\

\textit{Executing research} & 
Performing detailed calculations or simulations geared toward answering the research question  \\

\textit{Drawing conclusions/revising} & 
Interpreting the results of calculations, identifying inconsistencies with known observations or physical constraints, modifying approach to the problem  \\

\textit{Sharing results} & 
Publishing results of research in journals, giving talks, going to conferences, incorporating research into classroom activities \\

\hline \hline
\end{tabularx}
\caption{\label{tab:timecodes} Code descriptions generated to identify the approximate sub-processes involved in theorists' overall problem-solving processes.  Sub-processes were identified independently and were found to largely agree with prior results in \cite{park2009analysis}.}
\end{table}

We begin by describing results from the initial steps of our analysis during which we classified the sub-processes involved in theorists' overall problem-solving processes.  Each of the sub-processes were categorized based on their end goals, so identifying them allows us to better understand the context and purpose of theorists' analogies and assumptions. Indeed, we will later examine the nature of assumptions across these sub-processes. Hence, although the primary focus of this study is to characterize the nature of assumption and analogy use, we believe it is useful to provide a short description of these initial results here.  The sub-processes also had several entry conditions, since certain activities were unlikely or impossible to occur if other requisite activities had not been completed.  For example, idea generation almost always took place earlier in time than others, since analysis cannot be done without first identifying a research problem.  Thus, sub-processes were also loosely coupled with the time order of events in a theorist's overall process.

Over the course of coding five of the 11 interviews with the action/cue/sensemaking framework described in Section \ref{sec:Method}, we identified 663 actions and 326 instances of sensemaking, with 472 cues explicitly referenced as prompting those actions and sensemaking events.  Hence, we identified approximately 200 instances of actions or sensemaking per interviewee, with about half of those including an associated cue.  These were used to generate process diagrams, such as the one presented in Figure \ref{fig:callout}.  This diagram represents Dr. Dunn's research process and illustrates numerous examples of the action/cue/sensemaking events that we observed.


One of the standout features of these diagrams is the nonlinear nature of the theorists' research.  Although there generally appears to be a main line of progression from the beginning to end of the diagrams, there are always numerous offshoots representing activities such as the evaluation of new results or working through unforeseen challenges.  Visually, arrows connecting the action/cue/sensemaking bubbles often make loops representing iterative activities during which theorists jump between different points in their overall process as they refine their goals and progress toward a satisfactory research product.  The nonsequential nature of the theorists' processes aligns with previous results on the nature of authentic decision-making by STEM researchers \cite{price2021detailed}. 

Construction of these process diagrams allowed us to split theorists' research projects into several sub-processes that we identified based on the goals of the activities taking place within those sub-processes.  We identified five general categories describing these sub-processes of theorists' research: idea generation, preliminary design/analysis, executing research, drawing conclusions/revising, and sharing results.  These are generally consistent with prior work identifying the processes involved in physicists' authentic research practices as described in the Background.  Short descriptions of the sub-processes that we identified are provided in Table \ref{tab:timecodes}.    

We classified excerpts as idea generation when a theorist was trying to recognize a potential new avenue of research.  Idea generation included activities such as reading journal articles, discussing with collaborators, or examining experimental data.  Once theorists chose a potential new research project, they typically engaged in various preliminary planning activities in order to determine whether the project is doable, generate ideas about how to move forward in the research and which methods to use, and occasionally running a series of quick simulations in order to guide analytical work.  Actions in this category did not involve detailed calculations and did not necessarily produce the research products that theorists ultimately include in their published work.  Activities that involved performing detailed calculations and simulations intended to answer the research question fell under the category of executing research.  We categorized events as drawing conclusions/revising that involved the theorist reflecting on results and making decisions about whether to modify their approach to the problem.  For example, a theorist might have recognized that their predictions did not match observed experimental values, meaning that some aspect of their work was likely incorrect.  Lastly, we categorized activities as sharing results when they involved telling others about their research outcomes.  Theorists shared results by publishing results of research in journals and attending conferences, among others.  We do not address this stage in subsequent results, since it was not part of theorists' main analysis and never involved discussion of assumptions or analogies.


\subsection{\label{sec:Assumptions}Assumptions}

\begin{table*}[]
\centering
\def\arraystretch{1.5}
\begin{tabular}{P{0.2\textwidth}P{0.4\textwidth}P{0.4\textwidth}}
\hline
\multicolumn{1}{l}{\textbf{Type}} &
\multicolumn{1}{c}{\textbf{Inclusion}} &\multicolumn{1}{c}{\textbf{Assumption Description}} \\ \hline \hline

\textit{Universal} & 

Refers to assumptions about the model defining the underlying space that will be populated by the model's constituent objects (e.g., particles, planets, cell structures).  An example of a universal assumption in a model of the Newtonian World would be that the particles and interactions take place in a three-dimensional Euclidean space.      &

$\cdot$ Cell is a box \linebreak
$\cdot$ Wiring of a device \linebreak
$\cdot$ 1 to $n$ waveguides \linebreak
$\cdot$ ``Tweak" experimental setup \linebreak
\\

\textit{Constituent objects} & 
Refers to choosing which things are in the model.  This includes the object of interest in the model as well as the things in that object's environment.  In a Newtonian World the constituent objects might be a set of point particles. &    

$\cdot$ New particle/interaction \linebreak
$\cdot$ 1 to 2 cell filaments \linebreak
$\cdot$ Add proteins to cell \linebreak
$\cdot$ Limit number of cell building blocks \linebreak
$\cdot$ No dark matter \linebreak
$\cdot$ Limit number of model parameters
\\

\textit{Properties} & 
Assumptions about an object's properties refer to attributes that can be ascribed to the constituent objects.  These can be either properties that do not change (e.g., in the Newtonian World, the physical property of mass), or changeable properties (e.g., in the Newtonian World, the velocity of a particle).  &

$\cdot$ Curl of field is nonzero\linebreak
$\cdot$ Choosing particle spins \linebreak
$\cdot$ Filaments are Legos \linebreak
$\cdot$ Filaments are 1 dimensional \linebreak
$\cdot$ Non-spherical matter distribution \linebreak
$\cdot$ 1 to 2 dimensional $B$-field lines \linebreak
$\cdot$ 2 to $n$ sided wormhole \linebreak
$\cdot$ Choose equal mass black holes \linebreak
$\cdot$ Add black hole rotation \linebreak
$\cdot$ Add wormhole rotation \linebreak
$\cdot$ Choosing parameter values to match experiment \linebreak
$\cdot$ Choosing mitochondria types \linebreak
$\cdot$ Add dynamics to mitochondria \linebreak
$\cdot$ Look at boundary values/limiting cases
  \\

\textit{Interactions} & 
Defines which constituent objects are allowed to influence each other, as well as the ways that they are allowed to interact.  For instance, in the Newtonian World, two point particles might be allowed to interact with each other and their interaction is governed by Newton's Law of Gravitation.  &

$\cdot$ Neglect/keep processes (generically) \linebreak
$\cdot$ Add turbulence \linebreak
$\cdot$ Choosing biological processes to model \linebreak
$\cdot$ Limit model processes \linebreak
$\cdot$ Add feedback to system \linebreak
$\cdot$ Add energy term to differential equation \linebreak
$\cdot$ Neglect/keep processes (generically) \linebreak
$\cdot$ $F_G \sim 1/r^{2+\alpha}$ \linebreak
$\cdot$ Allow off-axis angular momentum \linebreak
$\cdot$ Add spatial dependence to system \linebreak
$\cdot$ Ignore GR effects \linebreak
$\cdot$ Do mean field theory \linebreak
$\cdot$ Linearize equations
  \\

\hline \hline
\end{tabular}
\caption{\label{tab:assumptiontypes} Examples of the diverse assumptions that theorists described, organized according to a variation of \citeauthor{hestenes1993modeling}' \cite{hestenes1993modeling} description of conceptual models described in Section \ref{sec:Background}.}
\end{table*}

Our analysis reveals that theorists make and refine their assumptions throughout a project, frequently considering how altering their assumptions might impact their work.  Theorists do not relegate making assumptions to a single step in the problem-solving process.  Rather, making and evaluating the assumptions in their model is an ongoing process.  

We coded 128 instances of ``assumptions" across 11 interviews.  These instances included explicit references to model features that the theorist supposed to be true, as well as general comments about their attitudes toward assumptions in their project.  Some assumptions were referenced more than once per interview and were coded separately so that we could take into account each statement's unique context in the overall research process.  We did not analyze any assumptions that the theorists did not explicitly mention.  For example, if a theorist said that they assumed a black hole was rotating, we did not break that down into smaller assumptions (e.g., a black hole exists, GR effects are present).  

\subsubsection{\label{sec:assumptionkinds} What kinds of assumptions do theorists make?}

\begin{table*}[]
\centering
\def\arraystretch{1.2}
\begin{tabularx}{\textwidth}{Z Z Z Z Z}
\hline
\textbf{Type} & \multicolumn{4}{c}{\textbf{Sub-process}} \\ \hline
& \multicolumn{1}{c}{\begin{tabular}[c]{@{}c@{}}Idea\\ Generation\end{tabular}} & \multicolumn{1}{c}{\begin{tabular}[c]{@{}c@{}}Preliminary Design\\ \& Analysis\end{tabular}} & \multicolumn{1}{c}{\begin{tabular}[c]{@{}c@{}}Executing\\ Research Calculations\end{tabular}} & \multicolumn{1}{c}{\begin{tabular}[c]{@{}c@{}}Drawing Conclusions\\ \& Revising\end{tabular}} \\ 
\hline \hline

\textbf{Increase Complexity} &

\begin{footnotesize}
$\cdot$ Add wormhole rotation \linebreak
$\cdot$ Add black hole rotation \linebreak
$\cdot$ Nonzero curl of field \linebreak
$\cdot$ 1 to 2 cell filaments \linebreak
$\cdot$ 1 to $n$ waveguides \linebreak
$\cdot$ 2 to $n$ sided wormhole \linebreak
$\cdot$ New particle/interaction \linebreak
$\cdot$ Add turbulence \linebreak
$\cdot$ Add off-axis angular momentum \linebreak
$\cdot$ Add dynamics to mitochondria \linebreak
$\cdot$ Add spatial dependence to system
\end{footnotesize} &

\begin{footnotesize}
$\cdot$ $F_G \sim 1/r^{2+\alpha}$
\end{footnotesize} & 

\begin{footnotesize}
$\cdot$ Use a non-spherical matter distribution
\end{footnotesize} & 

\begin{footnotesize}
$\cdot$ 1 to 2 dimensional $B$-field lines \linebreak
$\cdot$ Add feedback to system \linebreak
$\cdot$ Add energy term to differential equation \linebreak
$\cdot$ Add back-reaction force
\end{footnotesize} \\ \hline

\textbf{Could be either} &
& 
\begin{footnotesize}
$\cdot$ Neglect/keep terms (generically) \linebreak
$\cdot$ Neglect/keep processes (generically) \linebreak
$\cdot$ Choosing biological processes to model \linebreak
$\cdot$ Choosing particle spins \linebreak
$\cdot$ Wiring an optical device \linebreak
$\cdot$ No dark matter
\end{footnotesize} & 

\begin{footnotesize}
$\cdot$ Neglect/keep terms (generically) \linebreak
$\cdot$ Neglect/keep processes (generically) \linebreak
$\cdot$ Choosing parameters to match experiment (e.g., particle interaction parameters, conditions within a cell) 
\end{footnotesize} & 

\begin{footnotesize}
$\cdot$ Neglect/keep terms (generically) \linebreak
$\cdot$ Neglect/keep processes (generically) 
\end{footnotesize} \\ \hline

\textbf{Decrease Complexity} &  
& 
\begin{footnotesize}
$\cdot$ Limit number of parameters (generically) \linebreak
$\cdot$ Limit number of cell building blocks \linebreak
$\cdot$ Cell is a box \linebreak
$\cdot$ Filaments are 1 dimensional \linebreak
$\cdot$ Filaments are Legos \linebreak
$\cdot$ Ignore GR effects \linebreak
$\cdot$ Choose simplest 4 mitochondria types \linebreak
$\cdot$ Order of magnitude calculations
\end{footnotesize} & 

\begin{footnotesize}
$\cdot$ Only look at boundary value/limiting cases \linebreak
$\cdot$ Make spatially dependent differential equation ordinary \linebreak
$\cdot$ Average over system (mean field theory) \linebreak
$\cdot$ Linearize equations
\end{footnotesize} & 

\begin{footnotesize}
$\cdot$ Choose equal mass black holes \linebreak
$\cdot$ ``Tweak" experimental setup to simplify
\end{footnotesize} \\ 

\hline \hline
\end{tabularx}
\caption{\label{tab:assumptioncomplexity} An alternate organizational scheme for the assumptions that we observed in the data that categorizes assumptions based on whether they are increasing or decreasing model complexity, and when they occurred in the overall research process.  Visually it is clear that adding complexity tends to occur in the idea generation phase.  This indicates one way that theorists generate new ideas is by generalizing previous models.  Assumptions that decrease complexity tend to occur while they are doing preliminary work or executing research, indicative of attempts to simplify models such that they become tractable.}
\end{table*}

To answer the first research question, which asked what kinds of assumptions theorists make in their research, we generated a ``taxonomy" of theorists' assumptions.  Since our interview protocol was focused on theorists' overall process of scientific inquiry rather than the explicit enumeration of their assumptions, we recognize that our taxonomy is not exhaustive.  However, it remains useful as a way to gain insight into the diverse assumptions that theorists use in their daily work.  We organized theorists' assumptions into four broad categories: Universal, Constituent Objects, Properties, and Interactions (outlined in Section \ref{sec:Background}).  The results of this classification procedure as well as definitions for these categories are presented in Table \ref{tab:assumptiontypes}.  Short descriptions of the assumptions that theorists made are included in the table.  


Assumptions classified as universal referred to aspects of theorists' models that defined the underlying space in which the rest of the model's constituent objects exist.  While designing her model for a cell, biophysicist Dr. Costa decided that treating a cell as a ``box" was sufficient to achieve her project goals. While some researchers might need to model a cell's boundary in great detail to answer their research questions, Dr. Costa's work allows her to simplify her simulations by treating the cell as a box in which cellular structures are created: ``My goal was to understand how things are created inside of a cell. For me, a cell is a box. It could be a square box, the box does not matter."  This choice defined the space Dr. Costa then populated with certain cell structures, and contrasts with other cell simulations that allow for interactions with cell's boundaries.  We also classified several statements relating to the experimental setup of a device as Universal.  For instance, Dr. Fisher's choice of adding an extra waveguide to a device defined the domain that the constituent objects (photons) are able to move.

Statements classified as constituent objects referred to assumptions defining the particular things that theorists chose to include in the model.  These objects could include the object of interest to the theorist or objects that exist in the environment.  One of Dr. Jackson's primary assumptions in his research project was the existence of a new particle to extend the standard model.  Working under this assumption, he sought to explain how experimentalists would be able to confirm such a particle's discovery: ``We asked, let's say, despite the challenges, you managed to produce one of these new partner particles at the Large Hadron Collider...if you end up producing all of these partners at the LHC, how do you know that what you produced is in fact something that has to do with [this mechanism]?"  Thus, the introduction of the new particle in his model played a central role in the research project.  

Along with the constituent objects, theorists made numerous assumptions regarding the properties of those objects.  These referred to the attributes, fixed or changeable, that the objects possessed.  This was the most common type of assumption that we observed, and included statements regarding whether objects moved or rotated, their dimensionality, and their shapes. Some of these properties were static; Dr. Costa enforced the condition that the filaments that she was modeling were one dimensional and could not change.  Meanwhile,  Dr. Bahl and Dr. Huang, both working in the realms of cosmology and astrophysics, made explicit reference to the ability of the wormholes and black holes in their models to rotate with non-zero values of angular momentum, as opposed to being static.  The mitochondria in Dr. Irvine's cell models were dynamic entities that could move and change depending on their environment.   

The ways in which the constituent objects are allowed to influence each other and thereby change their properties were classified as Interactions.  Assumptions about how objects interact were the second most common type of assumption that we observed, and included numerous statements about the \textit{processes} being modeled.  For instance, Dr. Khan referenced the addition of a new type of feedback loop that had not been included in his previous models.  Other theorists such as Dr. Agarwal specified general laws governing interactions in their work: ``It's basically Newton's law of gravity, instead of falling like one over $r$ squared, it would fall like one over $r$ to the two plus alpha, where alpha is very small."  In this case, Dr. Agarwal explicitly makes an assumption about the modified form of Newton's law of gravitation, which defines the interaction between particles.

We also found it insightful to organize assumptions based on whether they increased or decreased the model's complexity.  The results of this classification scheme are presented in Table \ref{tab:assumptioncomplexity}.  Assumptions that increased complexity were those that appeared to add new complications or add generalizability to the theorists' models.  These included adding new processes into their model, adding new movement such as rotation, or increasing the number of items being modeled simultaneously.  Assumptions that we classified as decreasing complexity were those that simplified the researcher’s model, potentially making it less generalizable.  These included choosing to ignore difficult-to-model physical processes, performing math calculations that average over nonlinear processes, or reducing the dimensionality of a problem.  Some statements about assumptions (e.g., choosing to neglect or keep certain terms in an equation) were ambiguous as to whether they added or took away complexity in the theorists' models, so were classified as ``Could be either."  On a second axis is the problem-solving sub-process during which the theorist made the assumption. 

Organizing assumptions this way foreshadows several results regarding the roles that assumptions played in theorists' work.  Visually it is clear in Table \ref{tab:assumptioncomplexity} that making decisions that add complexity to a model tend to occur in the idea generation phase.  This indicates that theorists often generate new ideas by recognizing aspects of problems that they can generalize by making more complicated assumptions.  Similarly, several assumptions that increase model complexity appear in the process of revising the research, indicating that reexamining and adjusting assumptions to make the model more complex is one of the ways theorists try to troubleshoot problems that they encounter. 

On the other hand, assumptions that decrease complexity in a theorist's model tend to occur while they are doing preliminary work or executing research.  This feature of the table broadly illustrates theorists trying to simplify their models to make them tractable, but not so simple that the researcher cannot make progress on their research problem.  As stated by Dr. Dunn, ``You have to be careful, you know, it's all in the assumptions. And so theoretical physics is very much about making effective assumptions.... You can still do self consistent work if your assumptions are not valid, but then they don't, they're not correctly matching the observing."  In this quote, ``effective" assumptions are those that allow Dr. Dunn to progress on his research problem while also explaining the phenomena under investigation.  

Moreover, we begin to see in Table \ref{tab:assumptioncomplexity} the multitude of ways that assumptions interacted with the mathematics that theorists used.  For instance, preliminary design decisions such as limiting the number of parameters in a model preemptively made models simpler, while mathematical decisions such as averaging over space were made in response to a difficult mathematical problem that the theorist needed to solve.  This theme is discussed further in the following section.

\subsubsection{\label{sec:assumptionroles}What roles do assumptions play in theorists' problem solving processes?}

Assumptions primarily played an important role while theorists pursued the following objectives: 
\begin{itemize}
    \item Setting their project direction and goals
    \item Establishing how their model interacts with mathematics
    \item Revising their model while troubleshooting
\end{itemize}

We discuss each of these in turn, elaborating on specific examples of each and identifying several sub-themes within these broader categories (see Table \ref{tab:assumptionroles} in the Appendix for a list of all codes).

\subparagraph{Setting project direction and goals} Our analysis reveals that the overall project direction and goals that theorists set for their projects can be a direct result of a specific assumption that they have made about their model.  

We found instances in $N=9$ interviews of theorists referring to the identification of an assumption that they were able to vary, which lead to the generation of an idea for an entirely new research project.  The assumption could be in their own work or the work of others.  For example, Dr. Agarwal generated a new research idea while reading a journal article about the experimental observation of a particular relation between the radial acceleration observed in a galaxy and the galaxy's baryonic matter content.  While reading the article, her previous knowledge of the subject allowed her to recognize that the model the authors invoked to explain the relation was only applicable for spherically symmetric galaxies for which the curl of the gravitational field was equal to zero.  It was this realization that led to a new project idea: ``And the project that we looked at was examining how that assumption about the gravitational field...would affect the [relation]. So we specifically did not make that assumption." 
Thus, recognition of this implicit assumption using her existing knowledge of the subject led her to the question of whether the same relation would hold for non-spherically symmetric galaxies (e.g., discs).

In another example, Dr. Bahl recognized that a paper by a different group of researchers who had drawn several conclusions about the nature of wormholes used an the assumption that the wormholes were static.  According to Dr. Bahl, ``And so with my student, what we did is we asked the question of, `What if we consider rotation? What if the wormholes are rotating? Will their opening be larger or will the opening be smaller?'"  In this quote we observe the importance of the theorist's ability to recognize a breakable assumption in another paper and then identify potential repercussions of a more generalized model.  Dr. Bahl continued, ``So I have a very concrete goal.  I want to know how the size of the wormhole is affected by rotation."   Identification of a breakable assumption that the researcher believes will allow for generalization of previous work therefore translates to statements of specific project goals.     


We also observed that a theorist's overall goals and interests can influence their willingness to make certain assumptions or approximations in their work.  This was explicitly discussed in $N=2$ interviews.  For example, while explaining his work on elucidating the origin of magnetic fields in astrophysical objects, Dr. Dunn described his project goals in relation to the simplifications he was willing to make in the project.  He said,
``Well, for a complicated system like the sun and the ordered magnetic field, it's a major contribution if you can give the concept of how the field actually works. So you give a scenario that's somewhat simplified, but enough physics to be reasonable, and then you put that forward as a sort of paradigm."  Thus, his goal to put forth a general concept explaining the generation of magnetic fields in astrophysical objects influenced his willingness to make simplifying assumptions, so long as the physics met his threshold for being ``reasonable."  Dr. Dunn was striving for a mechanistic understanding of a system. Other physicists (perhaps one doing numerical predictions) may have had a different goal in mind, and may not have made the same simplifying assumptions.

Similarly, Dr. Costa discussed her desire to make enough approximations to make her project ``just right" in balancing model accuracy and utility.  
``I don't like completely simulation projects where, which are not analytically tractable. I like things which are just in the right spot. I don't want things to be too simple. But I don't want things to be so complicated that like a simulation that runs for 32 days is the only way to answer that question."  Whereas some biophysicists may be interested in modeling detailed properties of their system with so few approximations that they are only accessible via computation, Dr. Costa's personal interests as a researcher drive her to make assumptions (e.g., ``a cell is a box") that sacrifice realism for tractability.  As indicated by Dr. Dunn, the benefit of working on a more simplified case is that the results may act as a ``paradigm" that is applicable to a wide variety of problems rather than one highly specific system.

\subparagraph{Establishing how model interacts with math} 
\textbf{\textit{Model influences math:}} 
Researchers often referred to instances in which particular assumptions about the model caused them to take certain math steps ($N=8$ interviews).  In this way, assumptions influenced the proper mathematics to apply to their problem. Dr. Irvine's project involved modeling the population dynamics of mitochondria to predict the initial conditions that lead to healthy or unhealthy outcomes for the cell.  From the outset, Dr. Irvine knew that she would use four coupled differential equations to describe her model system because she assumed the existence of precisely four types of mitochondria in the system.  She recalled, ``The reason you have four equations is because you have four independent variables... 
We decided that, okay, these are sort of the simplest four silos we can make for this mitochondria.  That's why four variables, and then these variables are, because mitochondria are dynamical entities and they can fuse together or fragment, these equations are also coupled."  The number of mitochondria types, along with their dynamical properties, meant that Dr. Irvine could mathematically model her system as a set of four coupled differential equations. In this case, decisions regarding the model preceded the choice of math.  

Other theorists described this phenomenon in their work as well.  Dr. Bahl's decision to allow the wormhole in her model to rotate caused the mathematics to become analytically intractable, necessitating the use of numerical methods.  Comparing her group's methods to those of previous researchers, she said  ``They had the propagator, everything could be calculated analytically. But introducing the rotation introduced a level of technical difficulty that couldn't be solved analytically anymore and so all our results are in terms of plots." Again, making a particular assumption influenced the math the theorist used.  However, whereas Dr. Irvine's assumption directly informed her subsequent mathematical steps, Dr. Bahl's decision to include rotation did not immediately tell her that numerical analysis would be needed.  The decision to utilize numerical analysis came later in the research process, but was nevertheless a result of generalizing the wormhole model to include rotation.

\textbf{\textit{Math feeds back into model:}} 
Theorists also noted how their choice of assumptions while carrying out certain mathematical processes caused their model to change ($N=6$).  For example, once Dr. Dunn had decided how he wanted to approach the problem of understanding how large scale objects in astrophysics get their magnetic fields, he was confronted with a number of difficult equations to solve and needed to choose which approximations would be appropriate.  He described this decision making process by saying ``The system is very nonlinear, but I have to make approximations. So I say, I do what's called averaging, mean field theory, I average over the turbulent motions and say, I'm going to simplify all that into something and have equations for the average properties of the system." 
By deciding to simplify the nonlinear system using mean field theory, Dr. Dunn decided to limit his model to describing the average properties of the system.  The approximation needed to make progress on the mathematics therefore influenced his model's applicability.

When asked how he identifies which approximations to make, Dr. Dunn stressed the importance of experience in solving similar equations:  ``It's based on experience and judgment... knowing what solutions look like, which terms cause what kind of effects.  So I know having solved detailed sets of equations before, I'm comfortable knowing that dropping certain terms of certain types under certain circumstances is okay."  Thus, theorists are keenly aware of the impact that mathematical approximations may have on their model, a skill that is able to be learned with experience.  


Although we identified a number of cases in which it was clear whether decisions regarding the theorist's model influenced their mathematics or vice versa, it was sometimes difficult to differentiate precisely which came first.  This difficulty illustrates the tight interplay between mathematics and assumptions in theorists' minds.  When describing how she came up with the set of differential equations that she used to model cell mitochondria, Dr. Irvine recalled, ``We also wanted to do the simplest thing, so we did not have any spatial dependence... 
which meant that we were going to write down ordinary differential equations."  The decision to eliminate spatial variation in her model preceded the process of writing down the actual differential equations.  However, it was her foresight that the mathematics would be more difficult if the model included spatial dependence that drove the decision.  Similarly, when we asked Dr. Agarwal when exactly she began doing mathematical operations in her project, she identified the point at which she started ``doing math" as the same point as when she identified the assumption in a separate paper about the curl of the gravitational field being equal to zero.  She stated, ``Those are almost inseparable. Because it was like, oh, here's this function, the curl of this function is not zero, but they're assuming it's zero."  Her use of the phrase ``inseparable" underscores the extent to which assumptions and mathematics interact.

\subparagraph{Revising model while troubleshooting} Lastly, we find that researchers, cued by recognition that part of their work is undesirable (e.g., does not match experimental observation), often reevaluate their assumptions to troubleshoot the problem ($N=6$ interviews).  Thus, evaluation of their assumptions becomes a way to overcome difficulties in their work.

Dr. Garcia and Dr. Khan's project involved providing the theoretical framework for a new type of laser.  As such, their project required them to collaborate closely with experimentalists who were building the laser while they made theoretical predictions about its operation.  Dr. Garcia described his thinking when his postdoc Dr. Khan informed him that that their theory's predictions did not match experimental data, saying ``Well, did we mess up the theory or did the experimentalists not do a good enough job? Should we be including other processes? While that is being worked out, you tend to lose sleep."  Dr. Garcia's comment about including other processes indicates that one of the ways he would troubleshoot the mismatched theory and experimental result would be to revise the approximations he had made regarding which features of the model to include.

Experts undertook a continual process of examining whether the approximations that they made were sufficient to explain the physical phenomenon they were trying to model.  In the midst of this ongoing procedure, Dr. Dunn described checking his assumptions as a means of correcting problems with his model: ``you reach a point where you think this is just not working. And then you think through it, and then you know, you do have some sleepless nights...
So you can find yourself often in cases, you say, `Have I made the right approximations? Oh my gosh. Am I gonna have to go back and include that thing that was so helpful to drop, that term that was so offensive?' But now I have to include it."

Occasionally, a change of assumptions resulted in a drastic change to a theorist's model during the revision process.  Dr. Dunn realized that theorizing a model for the generation of magnetic fields in large scale astrophysical objects would require a mechanism to saturate the field growth so that the field would not grow infinitely. As Dr. Dunn described,  ``...what that led to was my reinterpreting... thinking not of the magnetic field as [one-dimensional], but as a new paradigm where you think of the magnetic fields as [two-dimensional]."  Changing a key assumption about the nature of the magnetic field allowed Dr. Dunn to push toward his goal, and had significant implications for similar studies in astrophysics.  However, other adjustments were not as drastic.  For instance, Dr. Jackson's research involved making predictions for the experimental signatures that would provide evidence for new particles.  When discussing procedures that he uses to overcome obstacles in his work, he stated that 
``...you usually try to ask yourself, can I sort of preserve the spirit of the project but change, tweak the setup a little bit so that I will overcome this challenge and still achieve more or less what I wanted to achieve."  Dr. Jackson's depiction of revising his model assumptions as ``tweaking...a little bit" differs drastically from Dr. Dunn's entire ``reinterpretation" of magnetic field lines.  The differences in scale of assumptions that the theorists are willing to adjust relates back to the previous themes of how assumptions influence the researcher's project direction, as well as the model and mathematics used.  Recognition that certain assumptions are unsatisfactory leads to new ones that can either alter the entire project direction or have much subtler impacts.

\subsection{\label{sec:Analogy}Analogy}

Utilizing the definition and framework for analogy presented in Section \ref{sec:Background}, we coded 87 instances of ``analogy" across $N=11$ interviews.  We coded excerpts that clearly referred to instances of the researcher mapping one piece of knowledge (the ``base” or ``source”) onto a concept or problem that the researcher is trying to explain (the ``target”).  To be coded as Analogy, statements had to have a clearly identified base and target. 


Analogies classified as high structure/low surface are frequently associated with those used in science (see Background Section \ref{sec:Background}).  These are analogies in which the base and target are from distant domains and do not appear at first glance to be related, and we observed examples of this analogy type in $N=7$ interviews.  For instance, Dr. Agarwal recognized that the equations of magnetostatics could be used to describe the galactic dynamics in her model.  She describes these kinds of analogies as ``mathematical analogies...filed under the principle of same equation, same solution" meaning that ``those equations have the same solution even though they have very different physical manifestations."  Despite sharing no surface similarity (current moving through a wire shows no obvious similarities to a disc galaxy), they shared similar underlying mathematical structures.

Meanwhile, analogies classified as high structure/high surface included instances in which the base and target problems were nearly identical save for only a few differences.  These types of analogies demonstrate that, in addition to the similar mathematical structures, there are a whole set of ideas, system behaviors, and other intuitions that one can map from one situation to the next.  We observed this type of analogy in $N=7$ interviews as well.  When confronted with the problem of modeling a circuit element that Dr. Fisher recognized as being highly similar to others previously modeled in the quantum optics literature, he immediately knew which mathematical method to apply to the problem.  He described the method he used as ``standard operating procedure" because the problems were so similar.  In this case, not only were the problems mathematically and structurally similar, they were also similar on a surface level (both were problems involving models of quantum optical circuit elements).  

These two types of analogy involving high structural similarity comprised nearly all instances of analogy use that we observed.  The lone example coded as low structure/high surface was a result of Dr. Fisher initially believing that the systems he was comparing would share similar underlying structures, but finding out that they only shared surface similarity. Hence, the analogy was not useful.  Specifically, when Dr. Fisher he attempted to apply the same ``standard operating procedure" to another similar circuit element, he found that the slight differences in the problems caused drastic mathematical differences that the standard methods could not resolve.  

\subsubsection{\label{sec:analogywhy}Why do theorists use analogies in their work?}

Theoretical physicists described using analogies in their work for a variety of purposes and at several stages of the research process.  We found that theorists primarily used analogies in order to:

\begin{itemize}
    \item Identify new project ideas and refine existing ideas
    \item Overcome conceptual obstacles
\end{itemize}

Within the broad category of overcoming conceptual obstacles, theorists employed analogical reasoning to help navigate difficulties determining the proper mathematical tools to use, to better understand the physical characteristics of a system, to demonstrate the feasibility of a project, and to build intuition about their research problem overall.  

\subparagraph{Identify project ideas and refine existing ideas} Through our analysis we discovered that a researcher’s analogical reasoning allows them to generate new ideas for a project, or can cause them to reframe their existing research question.  We noted this occurrence in $N=6$ of our interviews.  

The phenomenon of discovering a new project via recognition of an analogy was evidenced most strongly by Dr. Garcia, who recognized the possibility of constructing a new laser based on mechanical vibrations via analogy with standard optical laser physics.  Leading up to his identification of the analogy, Dr. Garcia was working on a research project that did not have to do with laser physics and revolved around a nonlinear equation.  After struggling with this work, he put his thoughts about the nonlinear equation ``on the back burner"  and began collaborating on an unrelated project with a colleague.  He then recalled, ``I found that I didn't have the background knowledge at enough depth to follow this guy's paper. So I said, okay, let me pull up this textbook. 
Imagine my surprise, when on the second page of [the] chapter, I saw the same equation staring out at me."  Despite reading about a seemingly unrelated topic, Dr. Garcia recognized the similar mathematical structure between the two systems.  He continued, ``And so this was a textbook on laser physics. It was describing how a laser works. And the equation was about the energy of the laser. And I'm like, this must mean we can make a laser out of our system." This analogy therefore provided the spark for an entirely new project idea.  Moreover, Dr. Garcia identified the power of the mathematical analogy as the reason he was confident in moving forward with this new project idea, saying ``So as soon as I saw the analogy, I knew that we could bring over the entire apparatus of existing optical laser theoretical description to this new system."  


While this instance of analogy usage led to an entirely new project idea, we also observed that theorists were willing to reframe the scope of their project due to an analogy that they discovered during the course of the research.  As discussed previously, Dr. Agarwal's stated project goal was to examine the effect of breaking the curlfree assumption of the gravitational field, which involved using Python and Mathematica to calculate and plot the divergence and curl of vector fields.  Over the course of troubleshooting this code to generate the proper output, she ``discovered a magnetostatic analogy in the context of [this theory]" and, noting that ``there's many other papers in the literature of people discovering these analogies," the elucidation of the analogy became a focal point of the final paper alongside the original goal.

\subparagraph{Overcome conceptual obstacles} Theorists used analogy as a tool to progress in their research when confronted with conceptual difficulties ($N=9$).  Indeed, Dr. Erdogan stated that when he is stuck on a difficult problem, ``...the next stage is that you argue by analogy.  That is, when you have a difficult, complicated problem that you cannot get your head around, you look for a similar problem that has already been solved, or that is easier to solve, which has some similarities to it."  Issues that theorists resolved using analogy often involved ambiguity as to which mathematical approach should be taken in their problem, or trouble thinking through the physical characteristics of the system.  The obstacles facing theorists included both specific issues such as trying to figure out how to solve a particular integral, or broader challenges like deciding how to begin attacking a problem in the first place.  In all of these cases, using analogy helped theorists overcome their conceptual challenges and make progress in their research.


\textbf{\textit{Math approach:}}
When Dr. Fisher tried applying a ``standard operating procedure" to a new circuit element, the slight differences in the new circuit caused drastic mathematical differences that he could not resolve.  He spent two years trying to solve this problem, until making an analogy to a different area of physics allowed him to resolve the problem in a few days.  He recalled,  ``It wasn't working. It was just, the matrices were too hard to sort out... So I started to think about how the matrix elements got in the matrix in the first place. And I realized that they were just like, they were very similar to something called Feynman path integrals that come up in a totally different context." 
Once he saw this similarity, he ``knew the problem was solved." This quote exemplifies how theorists use analogy to overcome difficulties in determining the proper math approach to the problem.  By relating the matrix elements in his problem to a mathematically similar problem in a different context, he quickly solved a yearslong problem.  
Using analogy to overcome conceptual obstacles related to determining their mathematical approach to a problem was the most common ($N=7$) way that we observed researchers using analogies.

\textbf{\textit{Build intuition and understanding:}}
Dr. Agarwal indicated that she knew how to make certain mathematical decisions based on 
``an intuitive feel."  Many theorists cited their intuition as important when making decisions.  Our interviews indicated that theorists employed analogous reasoning as one of the ways they built intuition and broad-based understanding about their research project ($N=5$ interviews).  

This strategy was particularly evident for researchers whose projects originated with inspiration from previous work, such as projects stemming from making a new assumption about a system (see Section \ref{sec:assumptionroles}).  For instance, while pursuing her goal of including rotation in her wormhole model, Dr. Bahl stressed the importance of being able to reproduce the results of the previous research paper that did not include rotation.  ``That's the purpose also of reproducing the results of the previous people. I mean, it's not only making sure that you know the technicalities that's important, but it also kind of makes you internalize the project, more like you really understand it, and so then you can have some intuition about it."
By reproducing the previous research and making it ``like yours, like if you had done it," Dr. Bahl was able to use the previous research as a base of knowledge to understand the new problem.  


While we acknowledge that analogies categorized as helping theorists determine their mathematical approach likely also helped them ``build understanding" about their problems, we chose to categorize them separately.  Specifically, to be coded as ``Math approach" we required that the relationship between making an analogy and the theorist's subsequent math steps was explicit.  Excerpts coded as building intuition and understanding lacked a description of precisely how the analogy translated to a subsequent mathematical approach to solve the problem.   

\textbf{\textit{Demonstrating a project is doable:}}
The final category we identified was demonstrating that a project is doable.  This was the least commonly discussed way that theorists used analogy ($N=3$ interviews).  While discussing how she typically starts working on a new project, Dr. Huang identified a common practice of solving a slightly simpler system in order to prove to her and her group that the more difficult problem is doable, as well as to gain understanding.  She stated, ``You start with something that usually is a proof of concept, okay?... If you're trying to develop, for example, an algorithm, you want to start by doing a simple test, devise a simpler system that you think can prove that you can handle that technically."  Hence, generating an analogous system to the system of interest gives Dr. Huang confidence in her ability to make progress.  This confidence is important, since researchers often seek to minimize the time spent working on problems that do not efficiently advance their research agenda.



\subsubsection{\label{sec:analogyhow}How do theorists identify analogies?}

The examples in the previous sections hinted at the diversity of ways that theorists are able to identify analogies.  We noted two primary categories for the ways that theorists found analogies:

\begin{itemize}
    \item Unplanned recognition
    \item Seeking out analogies
\end{itemize}

Importantly for students, theorists did not exclusively rely on existing knowledge and happenstance to find analogies.  Rather, researchers also actively engaged in seeking out or constructing systems that they could use as a base of knowledge to aid in problem solving.  

\subparagraph{Unplanned recognition}  In many instances ($N=9$), researchers recalled identifying an analogy spontaneously. In these cases the theorist did not deliberately seek out a system similar to the one they were working on.  Rather, their prior knowledge and awareness of other systems allowed them to recognize an analogy.

Dr. Costa said that she immediately knew that she could make progress on her problem because it shared many similarities with previous problems that she had worked on in graduate school.  Being able to map the knowledge from those old problems onto the new problem informed her choice of which computational method to use.  She recalled, ``So, like, the minute the question was given to me, I knew what method I wanted to do for it because I was familiar with that method from my bachelor's and my master's."  Dr. Costa did not need to look for new information to make the analogy between some of her previous work and her current research.  She ``knew" the computational method to use because of the similarities with her previous experiences. Several analogies that we discussed in previous sections fell into this category as well, including Dr. Agarwal's discovery of a magnetostatic analogy and Dr. Fisher's recognition of analogy with Feynman path integrals.  In each of these cases, the researchers spontaneously discovered that their problem possessed similarities to a different problem.

We distinguished these instances of unplanned analogy recognition from others based on whether they were \textit{prompted by an external source}, such as a textbook or research article.  For instance, Dr. Garcia's discovery of an analogy between an optical laser was spontaneous (he described being ``surprised" when he noticed similar mathematics described both systems), and was prompted by seeing a particular equation in an optical physics textbook.
 

\subparagraph{Seeking out analogies} The previous examples are juxtaposed with theorists deliberately seeking out an analogous system.  In these instances, the source of knowledge being mapped was external to the researcher and needed to be sought out.  Contrasted with descriptions from some researchers about how they ``discovered" or ``realized" or ``knew" an analogous system, other theorists described how they  ``look for" and ``find" a useful analogy.  For example, Dr. Erdogan stated that he will ``look for a similar problem that has already been solved, or is easier to solve."  He later reiterated, ``You have to find a toy model, a simpler version of this problem, which is easier to solve and which has similarities, some analogies to the actual problem."   

Theorists used several approaches to seek out analogous resources.  Dr. Irvine remarked that she consults with other physicists who are working on similar mathematical problems: ``So when you get stuck, one way to get unstuck is basically... talk to other people who may be experts in that area, or who are solving similar kinds of mathematical problems, but in a different context."  Through discussion with her colleagues, Dr. Irvine hoped to find a similar problem to her own that would allow her to overcome her conceptual difficulty.  Meanwhile, Dr. Bahl described seeking out previous research papers to help her construct and solve a particular integral.  She said that although her integral was difficult, ``...it's not that you have to come up with this integral completely from scratch. So people have worked on these [integrals] in this type of space before so there is literature that you can go and consult and read and all that... you have to adapt it somehow for your particular case, and then try to optimize to do it in the easiest way possible."  


\subsubsection{\label{sec:analogylimits}Limits of analogies}
Although theorists often espoused the benefits of using analogies, they also noted the importance of recognizing the limits of an analogy's usefulness.  We found that $N=4$ interviewees discussed needing to understand which aspects of the base system are reasonable and useful to map onto the target system. 


Dr. Garcia noted that although the mathematical analogy he made strongly indicated that he would be able to construct a new type of laser out of his mechanical system, the fact that the two systems were so disparate meant that there were numerous ways that the analogy could fail.  After his initial excitement finding the analogy, he recalled needing to verify ``the broad aspects of the whole theoretical structure and how we would map it onto our system."  He needed to make sure that certain approximations in the underlying theory for the laser system would be applicable to the new system, and remarked that ``the final scientific judgment on that only comes after the post doc has worked out every single term for our model and shown me, and convinced me, that the terms we need to throw away in order for the analogy to work are actually negligible in reality and if you were to do the experiment in the lab."  Thus, we observe that a crucial aspect of theorists' analogy usage is knowing which parts of the base system are amenable to being mapped onto the problem of interest.  

While reflecting on this process, Dr. Garcia noted that reading physics literature has helped him to sharpen his skills in determining which analogies will most likely help him advance his research goals.  ``Knowing the limits of the analogy is also important. So reading about such things, keeping them in mind. In the literature... you can have somebody describe the dynamics of a ping pong ball in an air jet or somebody talk about 11 dimensional gravity and string theory and they're all using analogies. It's just the nature of which one's more complex, and keeping track of the literature and seeing which analogies typically work out and which don't."  This quote emphasizes the fact that some analogies do not work as originally intended, and knowing their limits is important. It also indicates that students and researchers can improve their ability to recognize analogies.  Still, even if the analogy does not elucidate the specific aspects of the problem that a student or researcher wishes to know more about, the process of making an analogy, attempting to map it onto the original system, and recognizing its limitations could provide useful insight and intuition about the original system.  Thus, the analogy still may have served a valuable but different purpose than intended.

\section{\label{sec:Discussion}Discussion and Conclusions}



\subsection{Connections to modeling}

We analyzed both assumptions and analogies in this paper for several reasons.  One was to emphasize the similarities in the ways that theorists use them (e.g., coming up with new project ideas, overcoming challenges in their projects).  However, we also chose to address both topics in one analysis in order to discuss how analogous reasoning should itself be considered a modeling activity, similar to but distinct from those typically addressed in the physics modeling literature.

As alluded to in the Background section, many articles describing model construction portray the modeling process as a ``bottom-up" approach.  For example, \citeauthor{hestenes1987toward}' \cite{hestenes1987toward} strategy for model development 
initially focuses on the micro aspects of the system and builds toward a full model, culminating in a validation stage in which the researcher assesses the reasonableness of the assumptions included in the model.  Similar to the final stage in \citeauthor{hestenes1987toward}' modeling cycle, we observed that theorists often needed to reconcile model assumptions with mismatched predictions.  This particular similarity implies a unified way of thinking about models, assumptions, and analogies, since researchers must also determine the assumptions that allow their analogy to function as intended. Researchers make assumptions about which features are appropriate to map from base to target in order to make an analogy work, while simultaneously considering how those assumptions will affect their analogy's applicability and usefulness. This process parallels researchers' need to determine the appropriate assumptions for their model to predict observed phenomena.  

In fact, some physics education researchers explicitly link the processes of making models and making analogies.  \citeauthor{etkina2006role} write that ``explanatory models are based on analogies," and that students should use analogies to simplify aspects of reality (e.g., ``a car is like a point particle") \cite{etkina2006role}.  Students must also interrogate the effect that this simplification has on their model.  We observed similar sentiment in our data. Dr. Garcia stated that once he established his analogy with an optical laser, ``you had to go through a set of exercises similar to but distinct enough from the textbook model that you have to be careful about what you're keeping, what you're neglecting, which approximations are appropriate."  In this way, analogies act as a kind of model for researchers.  Dr. Garcia strongly hinted at this relationship, saying that he thinks of the base and target systems in his analogies as ``two different mathematical models, some of whose predictions agree with each other."  
Hence, recognizing the limits of validity for a model is a key concept in both analogical reasoning and modeling.  Yet  our results show that analogies can be used in far more diverse ways that just simplifying reality, and should serve as a basis for expanding on the ways that students and instructors can consider leveraging analogies in the classroom.  Rather than meticulously building models from the ground up (e.g., \citeauthor{hestenes1987toward}' modeling cycle), analogy allows the researcher to short circuit the model building process by porting over the details for their new model, including predictions about typical behaviors and understanding of the role particular parameters play.  Once their tentative model is established via analogy, the theorists can proceed to determine which aspects of the analogy are useful and map assumptions into a new domain.  In this sense, analogy represents a ``top-down" approach to modeling rather than the typical ``bottom-up" approach.

\subsection{Teaching implications \& future research}

This project has helped to make visible several of the ``invisible" expert processes that theorists use. Having illustrated these processes, we may begin leveraging expert approaches to support student learning.  Doing so will afford students the opportunity to engage in expert-like practices and improve their problem-solving skills.  This is especially pertinent for upper-division undergraduates and graduate students seeking to become professional physicists themselves. Integration of more professional practice into upper level curricula would naturally contribute to these students' socialization into the field, both by improving their problem-solving skills and by demonstrating more effectively what it means to ``do physics."  Moreover, it is likely easier to implement these practices into upper division courses via long-term projects based on open-ended questions that more closely mimic authentic research experiences.  However, we believe that appropriately tailored curricular changes could benefit aspiring physicists at any level, even if those changes manifest themselves differently in lower level undergraduate classrooms.  Whether undergraduate physics majors ultimately decide to pursue a career in STEM or not, giving these students exposure to the ways that make professional physicists' problem-solving skills useful and different would be beneficial. 

Although proposing a detailed curriculum redesign is outside the scope of this discussion, we suggest several ideas to help instructors begin thinking about strategies to encourage students to engage in these expert-like practices.  We hope these ideas will inspire future education research studies on how to best translate the results of our analysis into impactful curricular change.  Furthermore, we encourage readers to consider how these results might contribute to generating novel tools for assessing students' problem-solving.

\subparagraph{Assumptions}

If assigned problems involve physical setups that are fully specified in the problem statement, students may be limited in their ability to use and think about assumptions in their work.  This represents a potential area in which existing problem solving strategies in the PER literature may be augmented to further support student development.  Many physics problems are structured to provide the problem's assumptions and leave the student responsible for identifying the appropriate concepts and physics principles to apply.  Hence, PER has largely focused on novice/expert differences in how concepts and principles are selected and applied to problems.  Yet modeling frameworks identify both principles and assumptions as the two key inputs for model creation, and our analysis illustrates the importance of assumptions in real-world physics projects \cite{zwickl2013process}.  Thus, failing to afford students the opportunity to engage in making and evaluating assumptions limits their exposure to a key expert-like problem-solving technique.

As an example, we observed that theorists revise assumptions while evaluating their solutions and troubleshooting problems.  However, many problems give students little room to make assumptions at all, since they are already incorporated into the problem statement.  Students are therefore unlikely to evaluate their solution for reasonableness by considering whether their problem assumptions are sensible.  They are more likely to assume that they simply made an error in their math.  This limits students' capacity to execute this expert-like process, which could negatively impact their long-term expectations of where errors occur.



Several examples of research-based problem-solving strategies include the five-step problem solving strategy developed at the University of Minnesota \cite{heller1992teaching, heller1992teaching1}, case study physics \cite{van1991overview}, and the four step WISE strategy \cite{wright1986wise}.  Some of these existing research-based curricular designs already emphasize assumptions to varying degrees \cite{heller1992teaching, brewe2008modeling}.  Still, in the context of this study's results surrounding assumption usage, one could imagine adding to such existing problem-solving frameworks to more explicitly and authentically integrate the process of making and evaluating assumptions.  Critically, this gives students more agency to generate models, as well as for their solution path to be more open-ended.  

Instructors could implement a variety of activities to help students gain experience with generating New Project Ideas (Section \ref{sec:assumptionroles}).  For instance, instructors could extend textbook exercises by asking students to consider the implicit assumptions made in the problem and come up with new questions to ask based on changing those assumptions.  In a typical projectile motion problem such as a baseball flying through the air, students might ask how the problem would be different if the projectile was not treated as a point particle or if air resistance was included.  In a Gauss' Law problem asking to find the electric field around a uniformly charged sphere, students might ask how the problem would change if the distribution was not really a perfect sphere.  Would this change increase or decrease the problem's complexity? What predictions would be changed?  The original problems would need to be well-documented and understood by the student so that they could build on that knowledge.  On a larger scale, ideas for longer problems such as end-of-semester projects could be generated by generalizing problems done throughout the semester.

Future research must determine how to present assumption usage as a useful strategy to facilitate its transfer to students. Ideas about how to bring attention to the ways that assumptions interact with mathematics could be incorporated into a variety of instructional contexts.  Students might examine textbook problems to determine which assumptions are critical in the construction of the model, as opposed to those needed for calculational convenience at a later step.  Consider the case of  deriving the equation of motion for a simple pendulum.  To set up the problem in a way that allows students to apply their conceptual understanding to the physical system, a number of implicit assumptions (e.g., ignoring the rotation of the earth, ignoring air resistance) are typically made.  Later, it is typical to make a small angle approximation.  While the small angle approximation is not necessary to set up the problem, it makes the math analytically tractable.  Students could be tasked with motivating the reasons for making the various assumptions, including the small angle approximation, and the impact that they have on the pendulum model's applicability.  Questions to explicitly discuss with students could include:  Were the assumptions necessary to set up the problem, or were they made to move past mathematical difficulties that arose?  How did we know that they wouldn't have an adverse effect on our model's ability to accurately predict the pendulum's motion?  What are we defining as an ``accurate" model of the pendulum in the first place?  And if we wanted to generalize the simple pendulum model, what methods might be useful to do that?  Instructors know the answers to these questions, but they must make them apparent to students.



\subparagraph{Analogy}

Similar to how assumptions may be integrated into existing problem solving strategies, encouraging students to utilize analogies could also be implemented into PER-based instructional strategies.  Previous research by \citeauthor{clement1993using} has demonstrated that it is possible to effectively guide students' prior knowledge by utilizing detailed discussions of analogies \cite{clement1993using}.  \citeauthor{clement1993using} showed that class discussions about the validity of analogies between a target problem and a well-understood example helped improve student understanding.  In our data, several theorists noted that making useful analogies is a skill that can be improved with practice.  Further research into how to best weave analogical reasoning into existing problem-solving strategies could therefore benefit students in a multitude of ways.    

Explicitly calling out analogies as a useful tool could help draw students' attention to their utility.  Problem-solving processes could prompt students to think about whether they know of any other problems that might help them solve their current problem.  To help students utilize this skill more often, instructors may strategically choose exercises that students can use later as guides for solving more complicated problems; indeed, interleaving worked examples with homework exercises is already a well-known method of teaching problem-solving \cite{sweller1985use, ward1990structuring, pashler2007organizing} and promotes analogical reasoning.  This also gives students a chance to practice determining the limits of analogies that they make with other problems.  

Instructors may also design exercises that prompt students to deliberately construct an analogy, as we observed theorists do when approaching difficult problems.  When teaching RLC circuits, rather than solving the differential equation for the circuit outright, students might be asked instead to solve it using analogy with another system (the damped harmonic oscillator) and determine which features of each system and their behaviors are appropriate to map onto the other.  Perhaps students could develop analogies between a wave on a string with two clamped ends and a particle in an infinite square well. Students could consider questions such as: Where does the analogy work? Where does it break down?  Do any productive insights come from the analogy?  Answering these questions would help students see the limits of the analogy's validity and which aspects of the base string system are useful to map onto the square well.  Students may make some of these connections anyway, but encouraging more explicit attention via analogies could help students realize when these comparisons are useful guides or lead them astray.

Other research could examine the ways that students already use assumptions and analogies while solving problems.  Using a Cognitive Task Analysis framework could allow future researchers to study the particular aspects of context-rich problems and textbook problems that cue students to think about assumptions or analogies.  Also, more work could be done on students' productive analogy use, including how students justify mapping a base to a target and which aspects of the base they utilize.  This could be done in a natural setting by having students solve homework problems and documenting their analogy usage.  If students were solving problems about a harmonic oscillator, how many ideas do students import to the new system (e.g, mass, spring constant, resonant frequency, etc.)? How far do they take the analogy to draw insight? Does it offer mechanistic insight, or just a calculational path forward?  Answering these questions would help to better allow instructors to integrate this practice into their curricula.  

Lastly, considering how to leverage studies on expert practice to create more authentic long-term projects for upper-division undergraduates and graduate students could prove invaluable.  Project-based courses designed to engage students in more authentic problem-solving situations could help provide them with both useful skills and better insight into what ``doing theoretical physics" looks like. Therefore, we must continue to consider how to assess student usage and understanding of expert practices. This could provide opportunities for new assessment types designed to apply more broadly to theoretical physics problems and projects.


\subparagraph{Limitations} 

Although our interviews with theoretical physicists yielded rich data about the ways that they used assumptions and analogies in their research, there are several limitations to our study.  Our interview protocol was designed to elicit details about the decisions and actions of theorists throughout a research project.  We did not ask specifically ask about assumptions and analogies.  A protocol designed around these topics likely would have produced a more exhaustive account of the ways theorists make analogies and assumptions.  However, our study benefited from allowing us to observe these decisions in context, which a different protocol may have missed.  Also, since our interviews were retrospective accounts, it is possible that interviewees omitted or forgot about relevant aspects of their projects.  Other sources of data such as direct observation, discussions about in-process studies, and use of artifacts including preliminary calculation notes and published papers could uncover new aspects of theorists' cognitive processes.  



Researchers could also explore the relation between this study and work regarding how analogy can be used as a pedagogical tool.  For instance, how does analogy use for communication (e.g., in pop science) relate to analogy use for understanding causal mechanisms in a research process? Are people good at both?  Future studies could explore the similarities and differences between using analogy for sensemaking and using analogy for communicating to someone with less content knowledge.

\begin{acknowledgements}
We would like to thank all of the theorists who participated in this study.  This work was supported by the National Science Foundation Award DGE-1846321.
\end{acknowledgements}

\section*{Appendix: Tables of Assumption and Analogy Codes}
Tables \ref{tab:assumptionroles} and \ref{tab:analogyroles} provide summaries of the codes we used to classify the roles of assumptions and analogies.

\begin{table*}[t]
\centering
\def\arraystretch{1.1}
\begin{tabularx}{\textwidth}{|Z|Z|Z|}

\multicolumn{1}{c}{\textbf{Role of Assumption (Frequency)}} & \multicolumn{1}{c}{\textbf{Subcategory (Frequency)}} & \multicolumn{1}{c}{\textbf{Example}} \\ \hline 
\textbf{\textit{Setting project direction and goals (44):}}
\linebreak \linebreak
Applied to instances when the researcher refers to the goals of their project related to an assumption that they have made about their model, including the identification of a potential new project idea to explore & 

\textbf{\textit{New project ideas (27):}} 
\linebreak \linebreak
Researcher identifies an assumption in their work or the work of others that leads to the researcher generating an idea for the direction of a new project & 
\textit{``And the project that we looked at was examining how that assumption about the gravitational field...would affect [this theory]. So we specifically did not make that assumption. We said, let's not assume that the curl of this gravitational field is necessarily zero."} \\ \cline{2-3} 

\textit{} & 
\textbf{\textit{Broad goals and research interests (6):}}
\linebreak \linebreak
Refers to discussions of general goals that a researcher holds for projects regarding their approach to making assumptions and simplifications. This might include their propensity to use cruder models or fewer parameters in their work due to their interests and goals. & 
\textit{``What I'm after as a theorist is the principles here... so you can think of this: how accurately is a theorist interested in getting the right answer? Well, for a complicated system like the sun and the ordered magnetic field, it's a major contribution if you can give the concept of how the field actually works. So you give a scenario that's somewhat simplified, but enough physics to be reasonable, and then you put that forward as a sort of paradigm."} \\ \hline

\textbf{\textit{Establishing their model's interaction with mathematics (35):}}
\linebreak
\linebreak
Researcher describes how assumptions about their model influence and are influenced by the math that they are using in their project & 

\textbf{\textit{Model influences math (19):}}
\linebreak \linebreak
Assumptions about the model determine the subsequent mathematical steps that the researcher chooses to take & 
\textit{``The reason you have four equations is because you have four independent variables in terms of what amount of these mitochondria were small and healthy, what amount were big and healthy, what amount were small and unhealthy, what amount were big and unhealthy. We decided that these are sort of the simplest four silos we can make for this mitochondria."} \\ \cline{2-3} 

\textit{} & 
\textbf{\textit{Math feeds back into model (16):}}
\linebreak \linebreak
Researcher describes when mathematical steps (e.g., a choice to ignore nonlinear terms in order to solve an equation) alter the assumptions the researcher is making about the model & 
\textit{``The system is very nonlinear, but I have to make approximations. So I say, I do what's called averaging, mean field theory, I average over the turbulent motions and say, I'm going to simplify all that into something and have equations for the average properties of the system. So I don't need to solve the details of every random parcel of gas. I make an average over the system and solve a simplified set of equations."} \\ \hline

\textbf{\textit{Revising their model while troubleshooting (15):}}
\linebreak \linebreak
Refers to instances in which the researcher, cued by recognition that part of their work is undesirable (e.g., does not match experimental observation), reevaluates their assumptions &  & \textit{``...and he comes back and he says, it's not matching the data. Well, did we mess up the theory or did the experimentalists not do a good enough job? Should we be including other processes? While that is being worked out, you tend to lose sleep."} \\ \hline

\end{tabularx}
\caption{\label{tab:assumptionroles} Code descriptions and representative examples for assumptions.}
\end{table*}

\begin{table*}[t]
\centering
\def\arraystretch{1.1}
\begin{tabularx}{\textwidth}{|Z|Z|Z|}

\multicolumn{1}{c}{\textbf{Analogy: Why and How (Frequency)}} & \multicolumn{1}{c}{\textbf{Subcategory (Frequency)}} & \multicolumn{1}{c}{\textbf{Example}} \\ \hline

\textbf{\textit{Why - New project ideas and defining goals (19):}} 
\linebreak \linebreak
The researcher’s analogical reasoning allows them to generate a new idea for a project, or causes them to reframe their existing research question. & \textit{} & 
\textit{``Imagine my surprise when on the second page of that chapter, I saw the same equation staring out at me.  And so this was a textbook on laser physics.  It was describing how a laser works.  And the equation was about the energy of the laser.  And I’m like, this must mean we can make a laser out of our system."} \\

\textit{} & \textit{} &  \\ \hline

\textbf{\textit{Why - Overcoming conceptual challenges (51):}} 
\linebreak \linebreak Theorists use analogy to help them move past an issue in their project.  Issues often include ambiguity as to which mathematical approach should be taken in the  researcher’s problem, or trouble thinking through the physical characteristics of the system. &

\textbf{\textit{Math approach (29):}}
\linebreak \linebreak
Indicates the researcher was using analogy to overcome a conceptual obstacle related to determining the math approach to take in a problem, either in determining the proper math to apply to a problem or in trying to implement the proper math steps. & 
\textit{``And I realized that they were just like, they were very similar to something called Feynman path integrals that come up in a totally different context. But as it turned out, if you apply the same kind of reasoning to linear optical systems, the problem became super easy to solve, like, literally that day."} \\ \cline{2-3} 
\textit{} & 


\textbf{\textit{Build intuition and understanding (15):}} 
\linebreak \linebreak 
The theorist describes using an analogous resource in order to broadly improve their overall insight into their problem, rather than to resolve a particular difficulty. & 
\textit{``Yeah, reading and trying to understand very well with the other people have done... Essentially, you have to do it again. You have to do the results, reproduce the results of those previous guys from scratch, right."} \\ \cline{2-3} & 

\textbf{\textit{Demonstrate project is doable (4):}} 
\linebreak \linebreak 
Analogy is used in order to determine whether the researcher believes they are able to successfully make progress on the current research problem. & 
\textit{``If you're trying to develop, for example, an algorithm, you want to start by doing a simple test. So, you try to devise a simpler system that you think can prove that you can handle that technically."} \\ \hline

\textbf{\textit{How - Seeking out analogy (34):}} 
\linebreak \linebreak 
The researcher deliberately sought out an analogous system or problem to their own.  Thus the base of the analogy is external to the researcher's existing knowledge. &

\textbf{\textit{Reproducing previous results (8):}} 
\linebreak \linebreak 
A sub-code of Seeking out analogy that occurs when the researcher performing the calculations of a previous paper in order to use that knowledge base as the source for an analogy to their current project. & 
\textit{``So people have worked on these propagators in this type of space before, so there is literature that you can go and consult and read all that, but then you have to adapt it somehow for your particular case, and then try to optimize to do it in the easiest way possible."} \\ \hline

\textbf{\textit{How - Unplanned recognition (35):}} 
\linebreak \linebreak 
Applied to uses of analogy when the researcher spontaneously identifies the base analogous system. This discovery may have been preceded by significant time working on a problem, but they were not actively trying to generate an analogy during that time.  &  
\textbf{\textit{Prompted by external source (10):}} 
\linebreak \linebreak 
A sub-code of Spontaneous recognition that occurs when the researcher describes seeing or hearing about a resource (e.g., a textbook, research article, equation) that cues them to identify an analogy.
& 
\textit{``Then the actual sort of technical work that we did was very textbook. It was basically obtain divergence and curl of a vector field. In the process of doing that, we discovered a mathematical analogy between this [field] and a magnetic field."} \\ \hline
\end{tabularx}
\caption{\label{tab:analogyroles} Code descriptions and representative examples for analogies.}
\end{table*}

\clearpage


\bibliography{ms.bib}

\end{document}